 \documentclass[final,5p,times,twocolumn]{elsarticle}

\usepackage{graphicx}
\usepackage{amssymb} 
\usepackage{lineno}
\usepackage{textcomp}

\usepackage{amsmath}
\usepackage{siunitx}
\usepackage{tabularx}
\usepackage{amsmath,algpseudocode}\usepackage{algcompatible}
\usepackage{amsthm}
\usepackage{graphicx}
\usepackage{lscape}
\usepackage{verbatim}
\usepackage{color,soul}
\usepackage{xcolor}
\usepackage{subfigure}
\usepackage{tabularx,ragged2e,booktabs,caption,array,multirow,multicol}
\usepackage{gensymb}
\usepackage{csquotes}
\usepackage{mathtools} 
\usepackage{amsthm}  
\usepackage{listings}
\usepackage{amssymb}
\usepackage{latexsym}
\usepackage{epsfig}
\usepackage{float}
\usepackage{xspace}
\usepackage{algorithmwh}
\usepackage{float}
\usepackage{cuted}

\usepackage{algorithm}
\usepackage{algpseudocode}

\journal{Computers \& Geoscience}

\newcommand{\mb}[1]{\mathbf{#1}}
 \newcommand{\bs}[1]{\boldsymbol{#1}}
\let\today\relax
\begin{document}

\begin{frontmatter}


\title{Bayeslands: A Bayesian inference approach for parameter uncertainty 
quantification in Badlands}

\author[ctds,grg]{Rohitash Chandra\corref{cor2} \corref{con2}}
\ead{rohitash.chandra@sydney.edu.au}
\author[grg]{Danial Azam}
\author[grg]{R. Dietmar M\"uller} 
\author[grg]{Tristan Salles}
\author[ctds,math]{Sally Cripps}
\cortext[cor1]{Principal corresponding author} 
\address[grg]{EarthByte Group, School of Geosciences, University of 
Sydney, NSW 2006\\Sydney, Australia}
\address[ctds]{Centre for Translational Data Science, University 
of Sydney, NSW 2006\\Sydney, Australia}

\cortext[con2]{Rohitash Chandra contributed by  software implementation and 
write up of the paper. Danial Azam contributed mainly through software 
development and experimentation. R. Dietmar M\"uller contributed by steering 
the 
project, planning of experiments and editing the paper. Tristan Salles 
contributed the  section on Badlands and provided software    design support. 
Sally Cripps contributed by writing the paper and providing mathematical 
definition of the Bayeslands framework.  }

\address[math]{School of Mathematics and Statistics, University 
of Sydney, NSW 2006\\Sydney, Australia}
\begin{abstract}

Bayesian inference provides a rigorous methodology for estimation and 
uncertainty quantification of parameters in geophysical forward models.  
Badlands (basin and landscape dynamics model) is a landscape evolution model 
that simulates topography development at various space and time scales.  
Badlands consists of a number of geophysical parameters that needs estimation 
with appropriate uncertainty quantification; given the observed present-day 
ground truth such as surface topography and the stratigraphy of sediment 
deposition through time.  The  inference of unknown parameters is challenging 
due to the scarcity of data, sensitivity of the parameter setting and 
complexity 
of the Badlands model.   In this paper, we take a Bayesian approach to provide 
inference using Markov chain Monte Carlo sampling  (MCMC). We present  
\textit{Bayeslands}; a Bayesian  framework for Badlands that fuses information 
obtained from complex forward models with observational data and prior 
knowledge. As a proof-of-concept, we consider a synthetic and real-world 
topography with two parameters for Bayeslands inference, namely precipitation 
and erodibility. The results of the experiments show that Bayeslands yields a 
promising distribution of the parameters. Moreover, we demonstrate the 
challenge 
in sampling irregular and multi-modal posterior distributions using a 
likelihood 
surface that has a  range of sub-optimal modes.  
 
\end{abstract}


\end{frontmatter}
 

\section{Introduction}
\label{S:1}

A new generation of Earth evolution models has recently emerged with the 
capability to link models for dynamic and isostatic topography through time 
\cite{flament2013review} with landscape evolution models 
\cite{Braun1997,coulthard2001landscape,Whipple2002,Tucker10,Salles15,
Campforts2017,Adams2017,pazzaglia2003landscape,chen2014landscape}. This provides a model for landscape evolution in 
response to surface uplift and subsidence over a large range of spatial scales, 
and track sediments from source to sink 
\cite{salles2016badlands,salles2017influence}.   Geophysical forward  models 
depend on uncertain initial and boundary conditions \cite{scott1990finite}; 
such 
as global sea level, spatial and temporal variations in precipitation, and rock 
erodibility \cite{godard2006numerical,rejman1998spatial}. These models are 
calibrated against ground truth data which include the present-day topography 
and river geometries, total sediment thickness, sediment stratigraphy 
(constraining the time-dependence of sedimentation) and sedimentation rates, 
the 
time-dependent rock exhumation history from thermochronology data, geologically 
mapped paleocoastlines and geological data that constraints sediment 
provenance. 

 Bayesian inference provides a framework for combining information from various 
sources to estimate and quantify uncertainty of an unknown parameter in a 
rigorous manner  \cite{chair1988distributed}. The diverse source of information 
includes expert opinion,  ground-truth data, and knowledge from  physical 
processes embedded in geophysical models 
\cite{notarnicola2001bayesian,moon1995information}.
 Markov Chain Monte Carlo sampling methods (MCMC) implement Bayesian inference 
to sample from a posterior probability distribution 
\cite{hastings1970monte,metropolis1953equation}.   In the past few decades, 
Bayesian methods   became   popular in 
geophysics\cite{malinverno2002parsimonious,mosegaard1995monte,
sambridge1999geophysical}, where the emphasis has shifted from optimization 
methods  \cite{sen2013global} to inference due to the need for rigorous 
uncertainty quantification given sparse and  incomplete data 
\cite{gallagher2009markov}. The application of Bayesian inference via  MCMC 
methods in Earth science  has been demonstrated in a number of papers, such as  
characterizing geochronological data that describe  the age of rocks and  
fossils   \cite{jasra2006bayesian},  modelling the effect of climate changes in 
land surface hydrology \cite{raje2012bayesian}, calibrating hydrologic models 
\cite{thyer2015bayesian}, flood frequency analysis \cite{gaal2010inclusion}, 
inferring sea-level and sediment supply from the stratigraphic record 
\cite{Charvinetal09}, and inferring groundwater contamination sources 
\cite{wang2013characterization}. However, in the context of landscape  
evolution 
models, to our knowledge there is no work that employs MCMC methods to fuse 
information given by complex forward models with observational data. 

 \textcolor{black}{Landscape evolution models (LEMs) are characterized by 
parameters that interact 
in a complicated fashion and  feature a high dimensional parameter space  given 
time dependent  parameters  that represent climate factors for thousands to 
millions of years \cite{salles2016badlands}. Basin and landscape dynamics 
(Badlands) is an unusual example of a landscape evolution model for not only 
simulating topography development through time, but also tracking sediments 
from 
source to sink \cite{salles2016badlands} with the capability to create 
synthetic 
basin stratigraphies. We focus on using the Badlands software here because, 
despite the general abundance of landscape modelling software, there is no 
other 
mature open-source software that allows the simultaneous modelling of erosion, 
sediment transport and sedimentation. Badlands models feature a number of 
unknown parameters which need to be estimated given incomplete and sparse 
observational datasets which remains a major challenge  in the field. The 
challenge is in estimating the unknown parameters given  model complexity, lack 
of gradient information to construct efficient proposals, and computationally  
expensive models.  The posterior distribution of the  parameters in these 
models 
can  feature multiple modes and discontinuities which create further 
challenges. 
These have hindered the uptake of Bayesian inference methods for landscape 
evolution models \cite{salles2016badlands,salles2017influence}. }

\textcolor{black}{This paper presents Bayeslands, a  framework for inference and 
uncertainty 
quantification in the Badlands model for basin and landscape evolution.     We 
evaluate the performance of Bayeslands in terms of prediction accuracy of the 
final elevation and sediment erosion/deposition over time. Moreover, we   
present a visualization of the likelihood surfaces of the respective problem 
for 
a better understanding of the posterior distribution. }


\section{Background and Related Work}

\subsection{Bayesian inference}

  Bayesian inference regarding  a quantity of interest  $\theta$, is made via 
the posterior distribution   denoted by $p(\theta|\mb D)$, where $\mb D$ 
denotes 
the data. This posterior probability is proportional to the product of the 
likelihood $p(\mb D |\theta)$ and the prior $p(\theta)$, i.e. $p(\theta|\mb 
D)\propto p(\mb D|\theta)p(\theta)$. The prior distribution $p(\theta)$,  
reflects existing knowledge or belief about $\theta$ gathered from previous 
research and expert opinion. This prior is then updated via the likelihood 
$p(\mbox{data}|\theta)$, as  data are acquired.  The  likelihood $p(\mb 
D|\theta)$ is the probability that data are observed given some value of 
$\theta$.   The posterior probability quantifies the probability that $\theta$ 
takes on a particular value (if $\theta$ is a discrete) or a range of values 
(if 
$\theta$ is continuous).  The  posterior distribution, $p(\theta|\mb D)$, is 
rarely available in closed form, hence sampling based methods  such as MCMC  
are 
used for approximation.

 MCMC methods use the Metropolis-Hastings  algorithm to obtain draws of 
$\bs\theta$   from some proposal distribution. The draws of $\bs\theta$  are 
then accepted with a probability which ensures that the Markov chain satisfies 
the detailed balance condition \cite{tierney1998note}. If the proposed values 
are not accepted, then the chain stays at its current value.   Under certain 
conditions, the  draws of $\bs\theta$ will  converge to draws from the 
stationary distribution $p(\bs\theta|D)$. 
 
  The proposals distributions can include  random-walk proposals 
\cite{haario1999adaptive}, full conditional proposals  known as Gibbs sampling  
\citep{casella1992explaining}, and proposals which use gradients 
\cite{Neal2012}, such as the No U-Turn (NUTS)  sampler of Hamiltonian MCMC 
\cite{hoffman2014no,girolami2011riemann}.  The performance of these methods and 
their variants have been well studied in the   literature, see 
\citep{Robert2010} for a review.  Open-source Bayesian statistical packages 
have 
made the implementation of MCMC methods easier to facilitate  researchers  
without a background in Bayesian statistics  
\cite{fonnesbeck2015pymc,drummond2012bayesian}.


\subsection{Basin and landscape dynamics with Badlands model}

\textcolor{black}{Over the last decades, many numerical models have been proposed 
to simulate how 
 Earth's landscape  has evolved over geological time scales in response to 
different driving forces such as tectonics and  climatic variability 
\cite{Whipple2002,Tucker10,Salles15,Campforts2017,Adams2017}. These models 
combine empirical data and conceptual methods into a set of mathematical 
equations that can be used to reconstruct landscape evolution and associated 
sediment fluxes \cite{Howard1994,Hobley2011}. They are currently used in a 
variety of research fields including hydrology \cite{raje2012bayesian}, 
modelling sea level fluctuations, erosion, and sediment supply to basins and 
margins \cite{Charvinetal09}, the interplay between landscape degradation, 
vegetation and gully erosion  \cite{bastola2018role,gallen2018lithologic}.}

The Badlands model  simulates regional to continental sediment deposition and 
associated sedimentary basin architecture 
\cite{Howard1994,salles2017influence,salles2016badlands,salles2016b,
salles2017influence}  In its most simple formulation, the landscape  surface 
elevation changes in response to the interaction of three types of processes, 
(i) tectonic plate movement, (ii) diffusive processes and the associated 
smoothing effects,  and (iii) water flow and the associated  erosion. The 
change 
in elevation $z$  with respect to  time $t$ is given by
 
\begin{equation}
\frac{\partial z}{\partial t}=-\nabla \cdot  q_s + u
\label{eq:lem1}
\end{equation}
where, $u$ in $m\cdot yr^{-1}$ is a source term that represents tectonic 
uplift. 
The total downhill sediment flux $ q_s $ is defined by 

\begin{equation}
 q_s = q_r + q_d 
\label{eq:lem2}
\end{equation}

$q_s$ is the volumetric sediment flux per unit width ($m^2\cdot yr^{-1}$). 
$q_r$ 
 represents transport by fluvial system and $q_d$ hillslope processes both in 
$m^2\cdot yr^{-1}$.\\

\subsubsection{Fluvial system}

\textit{Badlands} uses a triangular irregular network (TIN) to solve the 
geomorphic equations presented in \cite{Braun1997}. The continuity equation is 
defined using a finite volume approach and relies on the method described in 
Tucker et al. \cite{Tucker2001}. To solve channel incision and landscape 
evolution, the algorithm follows the O(n)-efficient ordering method from Braun 
and Willett \cite{Braun2013}. This is based on a single-flow-direction (SFD) 
approximation assuming that water goes down the path of the steepest slope 
\cite{Callaghan1984}.

Several formulations of river incision have been proposed to account for long 
term evolution of fluvial system \cite{Tucker10,Chen14}. 
These formulations describe different erosional behaviours.  They include 
detachment-limited incision, which is governed by bed resistance to erosion, 
and 
transport-limited incision, which is governed by the capacity of the flow to 
transport sediment available on a river bed. Mathematical representation of the 
erosion process   is often assumed to follow a stream power law 
\cite{Hobley2017}. These relatively simple approaches have two main advantages. 
First, they have been shown to approximate the first order kinematics of 
landscape evolution across geologically relevant timescales 
($>$10\textsuperscript{4} years). Second, neither the details of long term 
catchment hydrology nor the complexity of sediment mobilisation dynamics are 
required. However, other formulations are sometimes necessary when addressing 
specific aspects of landscape evolution.

In this paper, our main objective centres around the estimation  with 
uncertainty quantification  of the parameters in Badlands. We use the default 
fluvial incision law available in \textit{Badlands} which is based on the 
detachment-limited stream power law, in which erosion rate $\dot{e}$ depends on 
drainage area $A$ ($m^2$), net precipitation $P$ ($m/yr$) and local slope $S$ 
which takes the form 
\begin{equation}
\dot{e}=\epsilon \, (P \cdot A)^m S^n
\label{eq:spl}
\end{equation}
where, $\epsilon$ is a dimensional coefficient describing the erodibility of  
channel bed as a function of rock strength, 
bed roughness and climate,   $m$ and $n$ are dimensionless positive constants. 
The default formulation assumes $m=0.5$ and $n=1$. Using this incision law, 
sediment deposition occurs solely in topographically closed depression and 
marine locations.

\subsubsection{Hillslope processes}

Along hillslopes, we state that the flux of sediment is proportional to the 
gradient of topography and a linear diffusion law commonly referred to as soil 
creep is used \cite{Tucker10,Salles15}. This is formulated as a diffusion 
equation
\begin{equation}
\frac{\partial z}{\partial t}= \kappa \nabla^2 z
\label{eq:hlinear}
\end{equation}
in which, $\kappa$ is the diffusion coefficient and can be defined with 
different values for the marine and land environments. It encapsulates, in a 
simple formulation, a variety of processes operating over short ranges on the 
superficial veneer of soils and sediments. $\kappa$ varies as a function of 
substrate, lithology, soil depth, climate and biological activity.

\section{Materials and Methods}
\subsection{Creation of Synthetic Data}

\begin{figure*}[htbp!]
  \begin{center}
    \begin{tabular}{cc} 
      \subfigure[Cr initial 
topography]{\includegraphics[width=90mm]{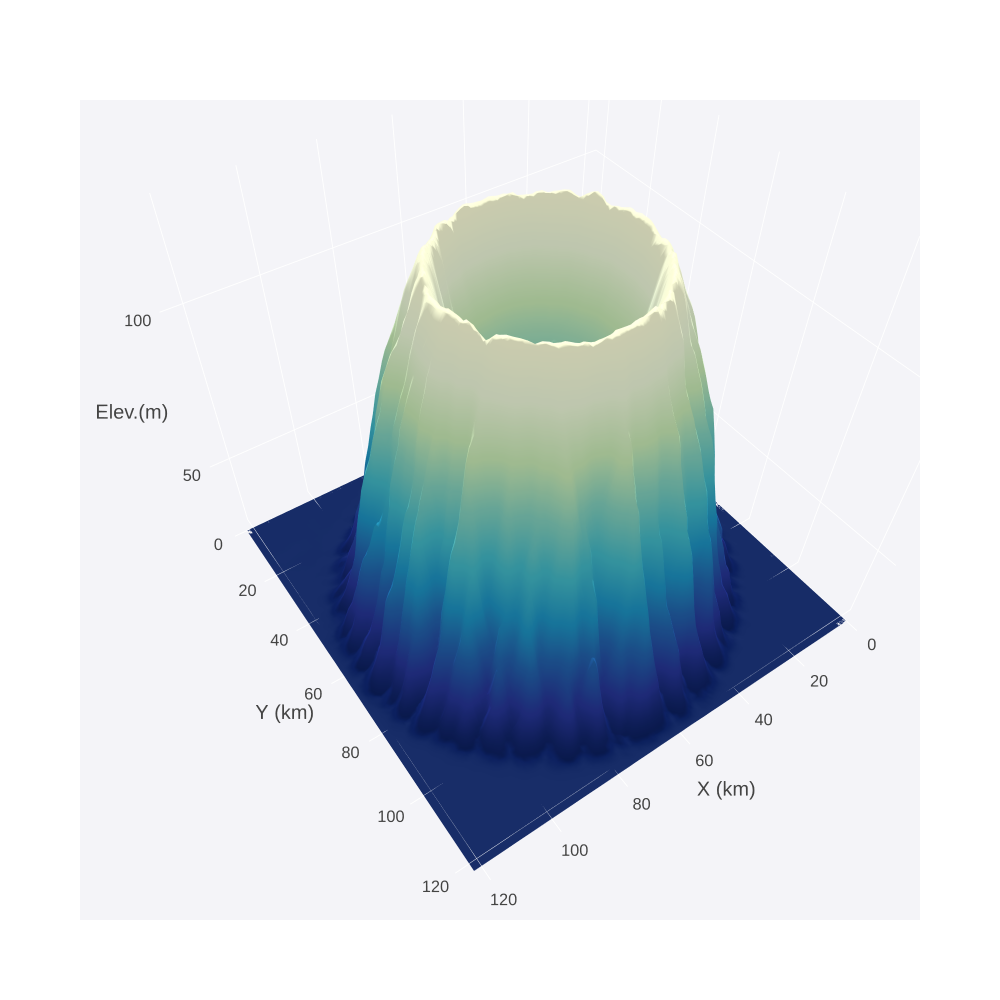}}
      \subfigure[Cr synthetic ground-truth 
topography]{\includegraphics[width=90mm]{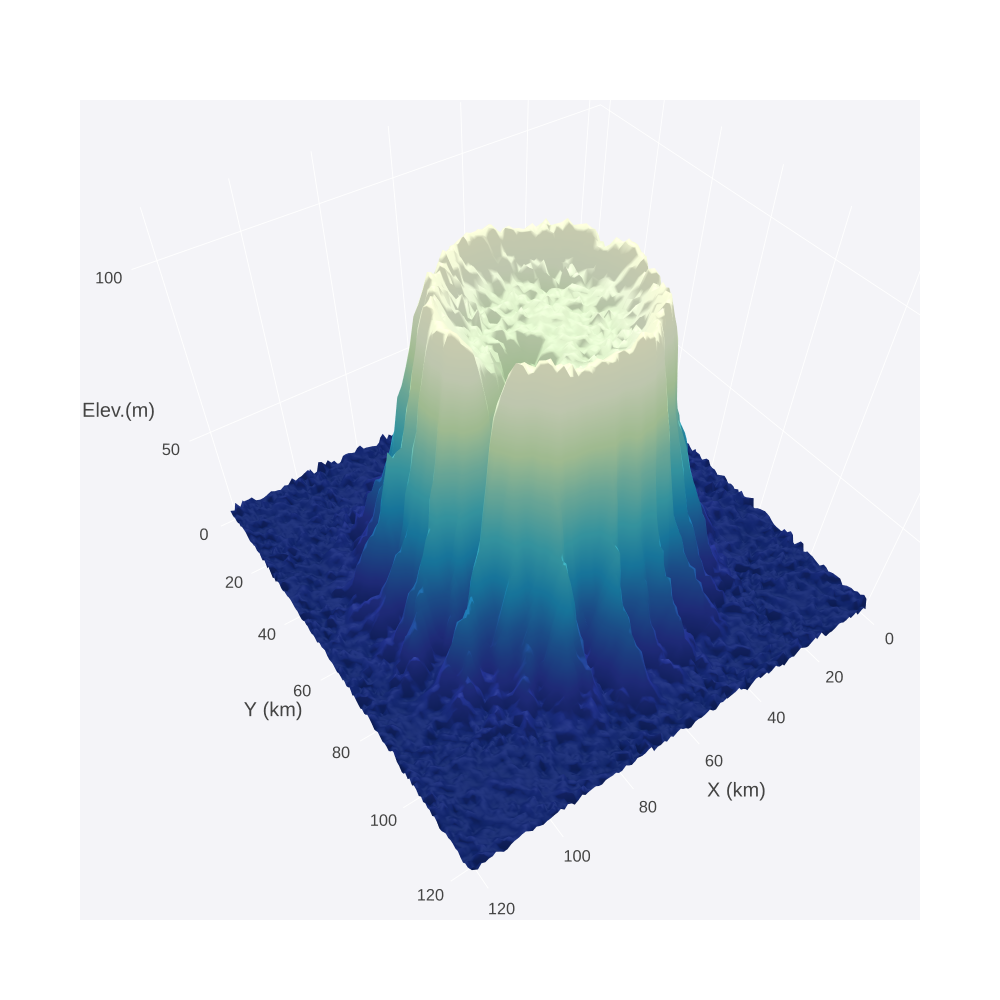}}\\  
     
    \end{tabular}
    \caption{Cr: Initial (panel a) and eroded ground-truth topography (panel b) 
after 50,000 years. }
 \label{fig:craterdata}
  \end{center}
\end{figure*}

\begin{figure*}[ht!]
  \begin{center}
      \subfigure
      {\includegraphics[width=100mm]{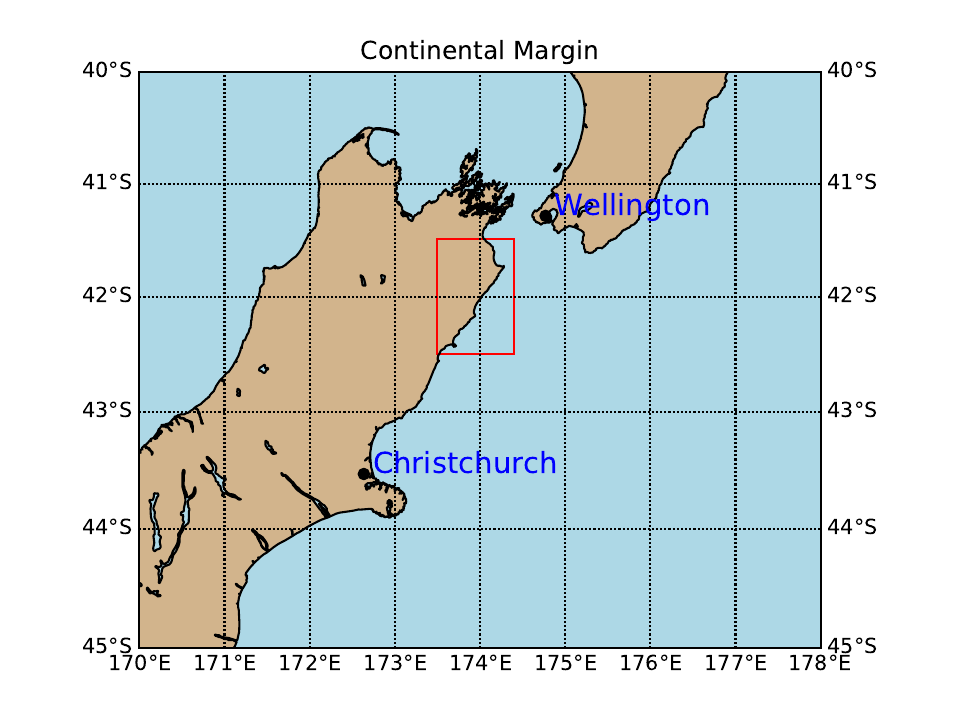}}
    \caption{Map showing the region (highlighted  in red) for the 
Continental-Margin problem selected from  the South Island of New Zealand.  
    }   
 \label{fig:cm-map}
  \end{center}
\end{figure*}

\begin{figure*}[htbp!]
  \begin{center}
    \begin{tabular}{cc} 
      \subfigure[CM initial 
topography]{\includegraphics[width=90mm]{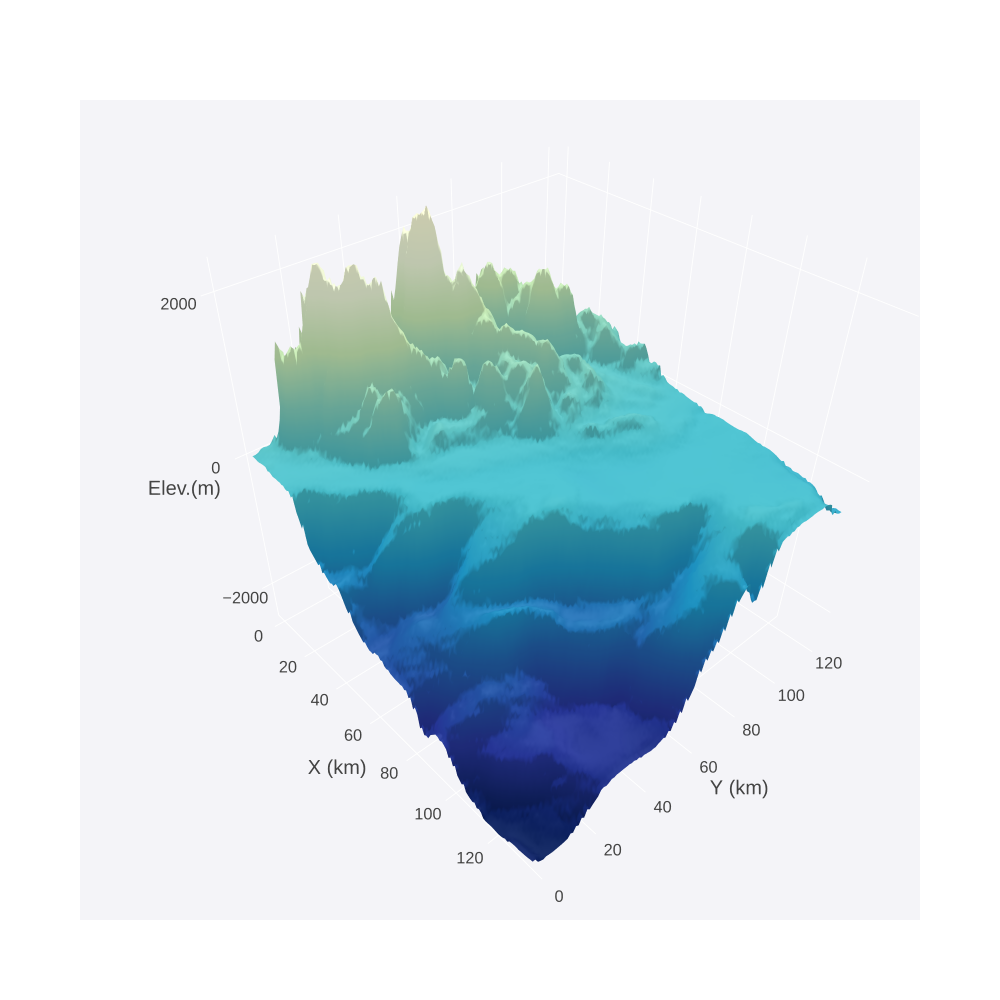}}
      \subfigure[CM  ground-truth 
topography]{\includegraphics[width=90mm]{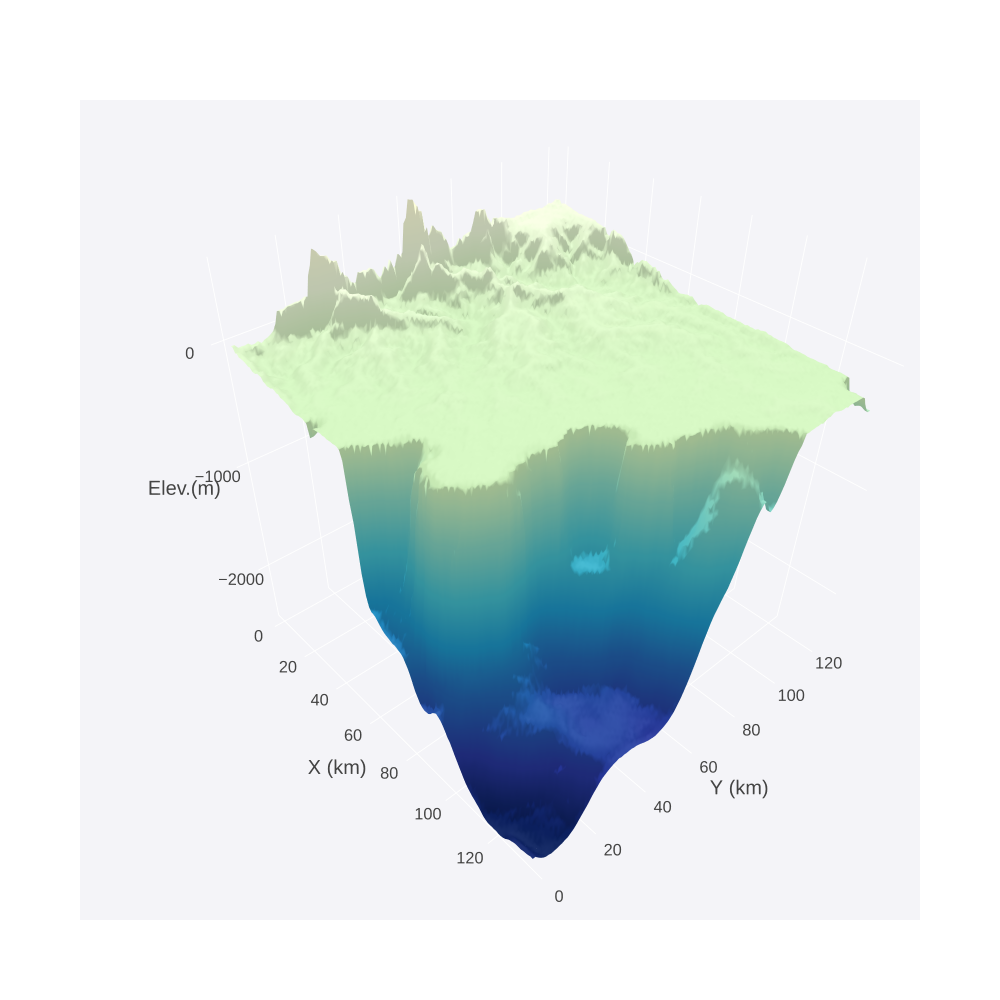}}\\  
      
    \end{tabular}
    \caption{CM: Initial and eroded ground-truth topography   after 1000,000 
years. }
 \label{fig:etopodata}
  \end{center}
\end{figure*}

\begin{figure*}[htbp!]
  \begin{center}
    \begin{tabular}{cc} 
      \subfigure[250,000 
years]{\includegraphics[width=80mm]{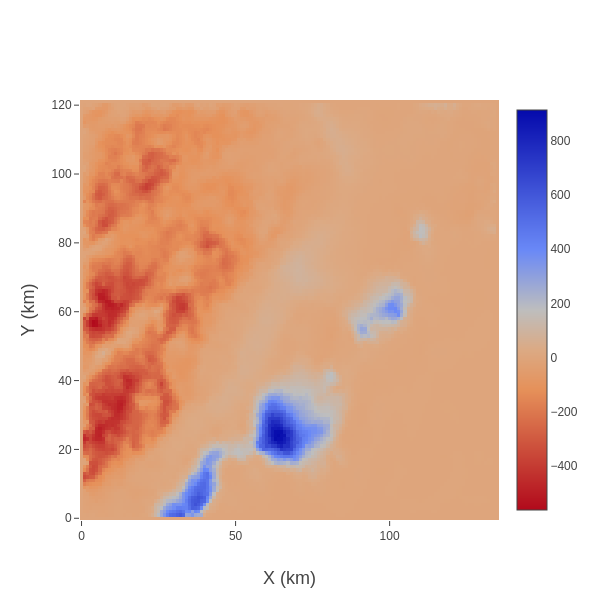}} 
\subfigure[500, 000 
years]{\includegraphics[width=80mm]{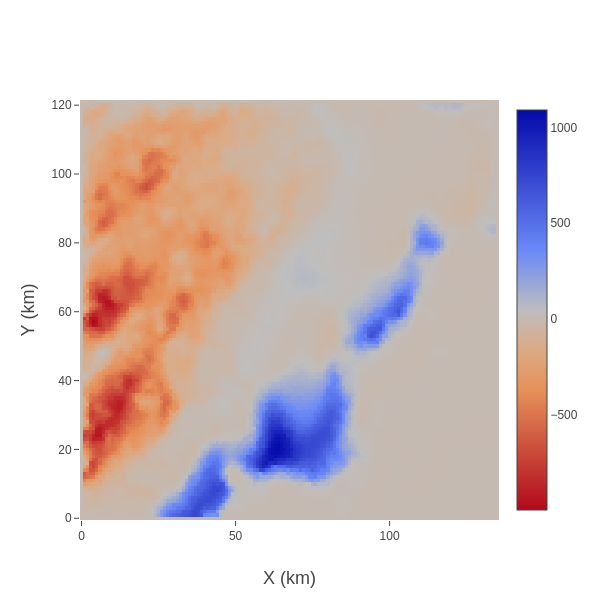}} \\
\subfigure[750, 000 
years]{\includegraphics[width=80mm]{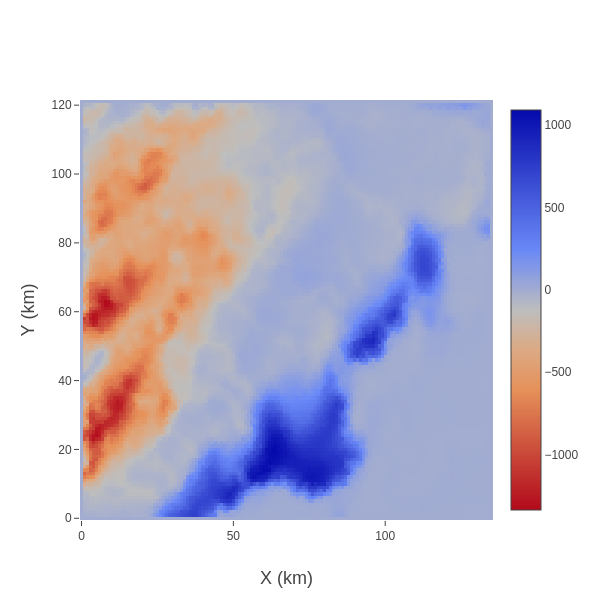}}  
\subfigure[1,000,000 years]{\includegraphics[width=80mm]{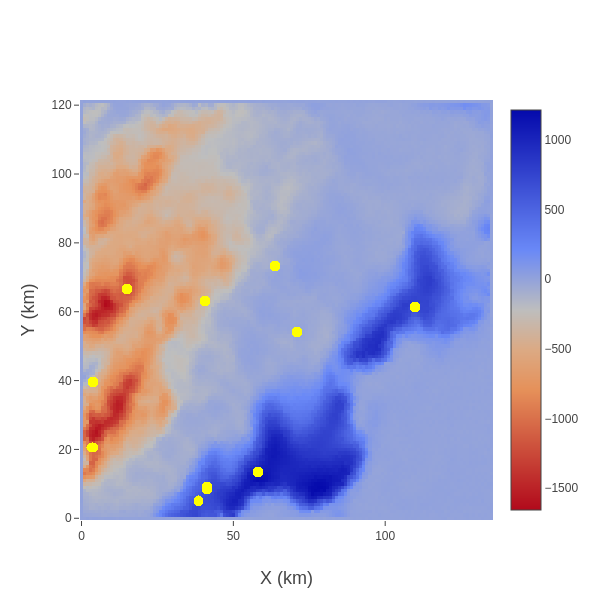}} 
\\
    \end{tabular}
    \caption{Sediment erosion/deposition for Continental-Margin   topography 
for 
4 selected time-frames, panels(a)~-~(d). Note that the erosion is negative 
while 
deposition is positive which is shown by the    legend   in meters. The yellow points 
mark the selected locations of the sediment data. }
 \label{fig:erodep_cm}
  \end{center}
\end{figure*}

In order to evaluate the performance of Bayeslands,  we consider two   
synthetic 
topographies that include the development of river systems, mountain ranges and 
sediment transport from source to sink. We refer to them as Crater (Cr)  and   
Continental-Margin (CM) and provide further details  as follows:
\begin{itemize}
\item   Cr : We simulate the geomorphological evolution over 50,000 years of a 
synthetic crater-type topography digital elevation model created by Badlands. 
The size of the   crater is    0.24 x 0.24 kilometers  squared. The topography 
is evaluated at a grid of  123 x 123 points (pixels). The resolution factor 
which defines the distance between two adjacent points in the grid. In this 
case, the resolution factor is   0.02 kilometer/point.  
Figure~\ref{fig:craterdata} shows the initial and the final topography  after 
50,000 years of evolution by Badlands. The final topography is used as the 
ground-truth topography.

\item   CM:  Using Badlands, we simulate the geomorphological evolution over 
1000, 000 years using a real  elevation  taken from present day South Island in 
New Zealand as shown in Figure \ref{fig:cm-map}. This region is represented by  
91 x 81 points that covers 136 by 123 kilometers squared;  hence, the 
resolution 
factor is 1.5 kilometers/point.   Figure~\ref{fig:etopodata} shows the initial 
and the final or ground-truth topography   by Badlands. Note that the CM 
topography is typically a 1 arc-minute global relief model of Earth's surface 
that integrates land topography and ocean bathymetry.  Figure 
\ref{fig:erodep_cm}   presents the sediment erosion/deposition for  selected 
time-frames with selected  locations used by the Bayeslands framework.

\end{itemize}

Table \ref{tab:truevalues} shows the set of  parameters that were used to 
create 
the  ground-truth topography described  above. We assume the  final 
topographies 
are observed in the present; by which we imply that the    Cr topography began 
evolution 50,000 years ago while the CM topography  began  evolution 1000,000 
years ago. Hence, the ground-truth or observed topography data ($\boldsymbol 
Y^{obs}$) refers to the  topography of present age, where $t=T $. The goal of 
Bayeslands is to sample and obtain a posterior distribution of the free 
parameters that governs  Badlands. The posterior distribution is used to make 
inference about the free parameters which includes prediction and uncertainty 
quantification.

\begin{table}[h!]
\small
\centering
 \begin{tabular}{c c c c c c } 
 \hline
Topography &   Time ( $t$ yrs) & $\rho$ (m/yrs) &  $\epsilon$ &     Run-time 
(s)\\  
 \hline 
 \hline
 Cr & 50,000 & 1.5 & 5.e-5   & 0.8  \\ 
 CM & 1000,000 & 1.5 & 5.e-6 &   1.1   \\
 
  \hline
 \end{tabular}
 
\caption{Values used for generation of synthetic ground-truth topographies (Cr 
and CM).   The simulated time that is represented by Badlands is given in years 
while the  run-time for executing a single Badlands model is given in seconds 
(s). }

\label{tab:truevalues}
\end{table}


 \subsection{Bayeslands Model and Priors}

\subsubsection{Model for  topography}

In order to implement  the model, we  take a probabilistic approach and assume 
that the observed elevation at time $t$, longitude $i$ and latitude $j$, is 
generated from a signal plus noise model 
\begin{equation}
y_{t,i,j}=g_{t,i,j}(\theta)+e_{t,i,j}
\label{eq_model1}
\end{equation}
where,   $y_{t,i,j}$ is the elevation, $g(\bs\theta)$ is the  signal that 
depends on the unknown 
parameters $\theta$, and $e_{i,j}\sim \mathcal{N}(0,\tau^2)$ is the noise. In 
this model, $\theta $ is equal to precipitation ($\rho$) and erodibility 
($\epsilon$).  The notation $N(\mu,\sigma^2)$ refers to the normal distribution 
with mean ($\mu$) and variance ($\sigma^2$).  The noise in Equation 
\eqref{eq_model1} reflects the fact that the elevation, at any point in time, 
will be affected by factors  other than those in the Badlands model.

Let $\mb Y_T=(\mb y_{T,.1},\ldots,\mb y_{T,.n})$; where, $\mb 
y_{T,.j}=(y_{T,1,j},\ldots,y_{T,n,j})'$ is the matrix of observed elevations 
across an $n\times n$  grid of latitude and longitude at time $t$.  We place an 
inverse gamma, $IG(\nu/2,2/\nu)$ prior on $\tau^2$ and integrate it out of the 
likelihood, denoted by $L_{Y_T}(\bs\theta)$, to give
\begin{equation}
L_{Y_T}(\bs\theta)\propto 
\prod_{j=1}^n\prod_{i=1}^n\left(1+\frac{(y_{T,i,j}-g_{T,i,j})^2}{\nu}
\right)^{-\frac{\nu+1}{2}}
\label{eq_like1}
\end{equation}

We note that the mean function $g(\theta)$ has no closed form representation in 
$\bs \theta$. Therefore, we cannot write down the likelihood as an explicit 
function of $\bs\theta$. However,  given  there exists a deterministic 
relationship between $\bs\theta$ and $g(\bs\theta)$, then 
$P(\mb Y_t|\bs\theta,\tau^2)=P(\mb Y_t|g(\bs\theta),\tau^2)$. We place uniform 
priors on precipitation  and erodibility, and the limits of the uniform 
distributions are given in Table~\ref{priorconf}

\begin{table}[htb!]
\centering
 \begin{tabular}{c c c  } 
 \hline \hline
Topography & $\rho$[min., max.]  & $\epsilon$  [min., max.]   \\  
 \hline 
 \hline
 Cr & [0.0-3.0]  & [3.e-5, 7.e-5]  \\ 
 CM & [0.0-3.0]  & [3.e-6, 7.e-6] \\
  \hline
 \end{tabular}
 
\caption{Prior range with minimum and maximum values  for precipitation 
($\phi$) 
and erodibility ($\epsilon$).}
 \label{priorconf} 
\end{table}

\subsubsection{Model for sediment erosion/deposition}

Due to the issue of multi-modality in complex models, there can be a  number of 
possible values of $\rho$ and $\epsilon$ which gives rise to the same 
topography. To constrain the number of possible values, we use information 
featured  in the  sediment erosion/deposition history. 


As stated previously, we only have ground-truth data of the 
landscape topography 
 in the present day denoted by $T $. The likelihood given by Equation 
\eqref{eq_like1} evaluates  the Badlands parameters $\bs\theta$ to ground-truth 
 
topography $y_{T ,i,j}$.  The estimates for surface topography   at previous  
timescales  are often  unavailable. However, sedimentary basins contain a 
record 
of   erosion and deposition which is available as total sediment thickness of 
as 
a stratigraphic sequence; i. e. sediment thickness through space and time, for 
$t<T $.  Therefore, the information from sedimentary deposits can be 
incorporated into the likelihood.
Let $z_{t,j}$ be the thickness of the sediment  at time $t$, at location $j$.  
We assume this deposition also depends upon $\bs \theta$ via a signal plus 
noise 
model whereby
\[
z_{t,j}=f_{t,j}(\bs\theta)+\eta_{t,j}
\]
where, $f(\bs\theta)$ is  the signal that represents  sediment thickness given 
by the  Badlands model, for a given value of $\bs\theta$ and $\eta \sim 
\mathcal{N}(0,\sigma^2)$ is the noise.   Analogous to the topography data, we 
integrate out $\sigma^2$ and the likelihood function for the sediment data, 
$L_Z(\bs\theta)$  is given by 

\begin{equation}
L_Z(\bs\theta)\propto 
\prod_{t=1}^{n_t}\prod_{j=1}^m\left(1+\frac{(z_{t,j}-f_{t,j})^2}{\nu}
\right)^{-\frac{\nu+1}{2}}
\label{eq_like2}
\end{equation}

where, $n_t$ is the number of  times that the sediment thickness is recorded 
and 
$m$ is the number of  points at which the sediment is measured.


 \subsection{Bayeslands framework}

We take a Bayesian approach and estimate the topography  via Badlands using the 
 
posterior mean of the selected parameters, $E(y^*_{T,i,j}|\mb Z_t,\mb Y_{T})$. 
The posterior mean is equal to
\begin{eqnarray*}
E(y^*_{T,i,j}|\mb Z,\mb Y_{T})&=&\int E(y^*_{T ij}|\mb 
Y_{T},\bs\theta)p(\bs\theta|\mb Z_t,\mb Y_{T})d\bs\theta\
\end{eqnarray*}

and an estimate of it is

\begin{eqnarray}
\hat{E}(y^*_{T,i,j}|\mb Z,\mb Y_{T})&=&\frac{1}{M}\sum_{j=1}^M 
E(y^*_{T,i,j}|\mb 
\mb Z_t,Y_{T},\bs\theta^{[j]})
\label{eqn_evalue}
\end{eqnarray}

where,  $\bs\theta^{[j]}\sim p(\bs\theta|\mb Y )$ is  obtained via MCMC in 
Bayeslands  as shown in  Algorithm ~\ref{alg:alg1}.  That is we draw each 
element of $\theta$ from a proposal distribution, $q(\theta|\theta^c),$ 
conditional on the current value of $\theta$, denoted by $\theta^c$.  The 
proposed value of $\theta^p$ is accepted with probability 
\begin{equation}
\alpha  = \min\left\{ 
1,\frac{L_{Y_T}(\bs\theta^{p})L_{Z}(\bs\theta^{p})}{L_{Y_T}(\bs\theta^{c})L_{Z}
(\bs\theta^{c})}\right\}
 \label{eqn_accept}
 \end{equation}
which ensures the convergence of the chain to its stationary distribution, 
$p(\theta|\mb Y,\mb Z)$.  Note that as $q(.)$ is symmetric and $p(\bs\theta)$ 
is 
the product of independent uniform priors, $\alpha$ reduces to the likelihood 
ratio. The inference regarding  the selected parameters are given via the joint 
posterior $p(\bs\theta|\mb Y ,\mb Z)$.

\begin{figure*}[htb!] 
\includegraphics[width=120mm]{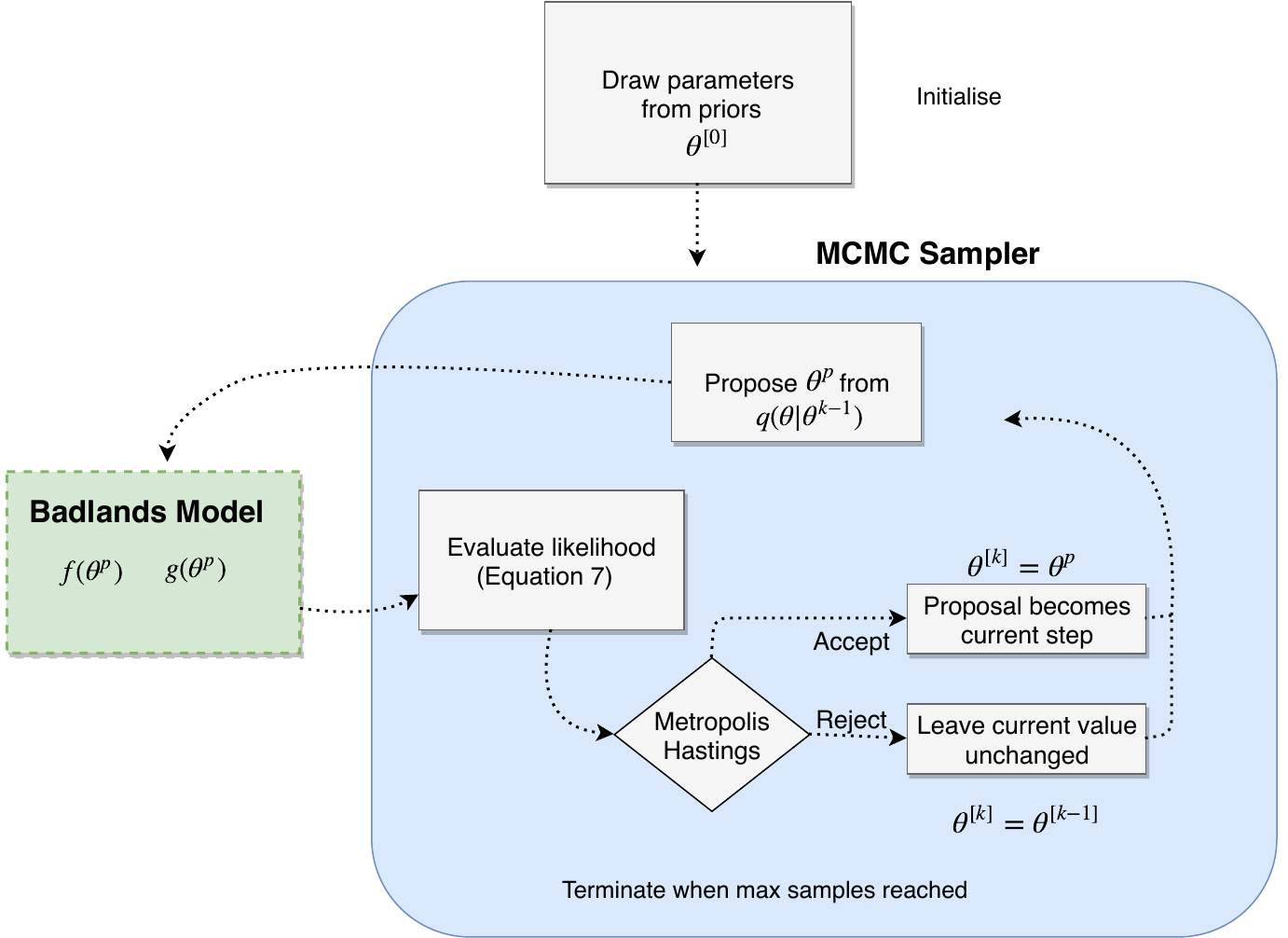} 
\centering
\caption{ \textit{Bayeslands} framework employing MCMC random-walk sampler and 
Badlands model. Further details are given in Algorithm \ref{alg:alg1}.  } 
\label{fig:Bayeslands}
\end{figure*}

Figure 
\ref{fig:Bayeslands} gives an overview of the Bayeslands framework that employs the MCMC sampler for the  Badldands model.  Algorithm 
\ref{alg:alg1}   begins by initializing values of  precipitation and erodibility 
by drawing from the  respective prior distribution for the selected parameters 
given in  Table~\ref{priorconf}. The algorithm then proceeds by  proposing new 
values of the parameter  (Step 1) from the   normal distribution as the 
proposal 
distribution, with mean $\bs\theta^{[i-1]}$ and     the step-size ($\phi_p$ and 
$\phi_\epsilon$ ). Conditional on these proposed values, Badlands is executed 
for the maximum time (eg. Cr=50,000 years) which produces outputs that include 
the successive sediment erosion/deposition and topographies (Step 2). The choice of a  random-walk proposal distribution was due to the 
unavailability 
of gradients for the selected parameters in the Badlands model. Then the 
likelihood given by Equation  \eqref{eq_like2} is evaluated by considering the 
final topography  and the successive erosion  deposition values at selected 
locations  (e.g. Figure \ref{fig:erodep_cm}).  Once the likelihood is computed, 
the Metropolis-Hasting criterion is used for determining whether to accept or 
reject the proposal (Step 3). If the proposal is accepted, the chain moves to 
this proposed value. If rejected, the chain stays at the current value (Step 
4). 
The process is repeated until the convergence criterion is met, which in this 
case is the maximum number of samples defined by the user. Note that the 
software package in Python along with data  is given online 
\footnote{Bayeslands: https://github.com/intelligentEarth/Bayeslands}.

\begin{algorithm}[htbp!]
 
\small
	\caption{\textit{Bayeslands} framework}
    \label{alg:alg1}
    
	 Initialize $\bs\theta=\bs\theta^{[0]}$  by drawing $\bs\theta^{[0]}$ 
from the prior $\bs\theta^{[0]}\sim p(\bs \theta)$ \\
	For $i=1:Samples$
	\begin{algorithmic}[1]
    
    \STATE Propose a value $\bs\theta^{[p]}|\bs\theta^{i-1}\sim 
q(\bs\theta^{[i-1]})$, where $q(.)$ is the proposal distribution. 
	\STATE Given  $\bs\theta^{[p]}$, execute   \textit{Badlands} and 
compute 
$g_{Tmax,i,j}\left(\bs\theta^{[p]}\right)$, and  
$f_{t,i,j}\left(\bs\theta^{[p]}\right)$  
	\STATE Calculate the acceptance probability $\alpha$, as given by 
Equation~\ref{eqn_accept}.
 
 \STATE Generate $u \sim U(0,1)$ and set $ \bs\theta^{[i]} = \bs\theta^{[p]}$ 
if 
$\alpha < u$. Otherwise, set $\bs\theta^{[i]} =\bs\theta^{[i-1]} $.
\end{algorithmic}
\end{algorithm}

\subsection{Experimental setting and evaluation}

\textcolor{black}{In order to  evaluate  the performance of Bayeslands on the  
problems (CM and 
Cr) presented previously,  we consider the following issues and metrics:
\begin{enumerate}
\item Visualization of the true posterior surface evaluated across the grid; 
\item Bayeslands MCMC sampling and posterior distributions; 
\item Prediction of the topography  and sediment erosion/deposition. 
\end{enumerate}}

\textcolor{black}{We present the metrics for evaluating Bayeslands using the 
Badlands outputs 
(predicted topography and sediment erosion/deposition).  We  highlight that 
although a series of topographies are generated by Badlands, only the final 
topography is used for evaluation, while selected sediment erosion/deposition 
is 
used. The root mean squared-error (RMSE) is used as the metrics for evaluation 
where  the final topography elevation (elev) and sediment erosion/deposition 
(sed) respectively are given by}

\begin{eqnarray*}
 \mbox{RMSE}_{elev} & = &\sqrt{\frac{1}{n\times m}\sum_{i=1}^n\sum_{j=1}^n 
\left(g(\hat{\theta}_{T,i,j}) -g_{T,i,j}(\theta)\right)^2}\\
 \mbox{RMSE}_{sed} & = & \sqrt{\frac{1}{n_t \times 
v}\sum_{t=1}^{n_t}\sum_{j=1}^m \left(f(\hat{\theta}_{t,j}) 
-f(\theta_{t,j})\right)^2}
  \end{eqnarray*}

 \textcolor{black}{where, $\hat{\theta}$ is an estimated value of $\theta$, 
chosen according to 
some criteria and $\theta$ is the true value  on which the ground truth 
topographies  and sediment thickness were based.  $f(.)$ and $g(.)$ represent 
the outputs of the  Badlands model, as defined earlier while $m$ and $n$ 
represent the size of the selected topography.  $v$ is the total number of 
selected points from   sediment erosion/deposition as shown in Figure  \ref{fig:erodep_cm} over the selected time frame, 
$n_t$. }
 
\textcolor{black}{We use the information about uniform priors for the respective 
parameters  from 
Table \ref{priorconf}. We  construct proposal distributions  using a covariance 
matrix, which is assumed to be diagonal with entries equal to the square of 
step-size ($\phi$). We performed several trial experiments to evaluate an 
optimal $\phi$  and use   $\phi_{\rho} = 0.03$ for precipitation for both cases 
(Cr and CM). In the case for erodibility, CM uses $\phi = 4.e-8$ while Cr uses 
$\phi_{\epsilon} = 4.e-7$.  We evaluate  Bayeslands for  selected number of  
samples with a 10\% burn-in period. Note that burn-in is considered as the 
initial sampling period before the draws in the chain are assumed to be from 
the 
invariant distribution, which in this case is the joint posterior. The 
different 
number of samples are used to check the convergence properties of the MCMC.}

The overall computation time taken for Bayeslands for each experiment is also 
reported. We use  an Intel Core i7-8700 Processor (Hexa-core, 3.2 Giga-Hertz) 
for all the experiments. Note that parallel computing  was not used in the 
implementation of  Bayeslands.

\section{  Results}

\subsection{Visualization of the true posterior  surface    }

\begin{figure*}[htbp!]
  \begin{center}
    \begin{tabular}{cc} 
      \subfigure[Cr true posterior  
surface]{\includegraphics[width=80mm]{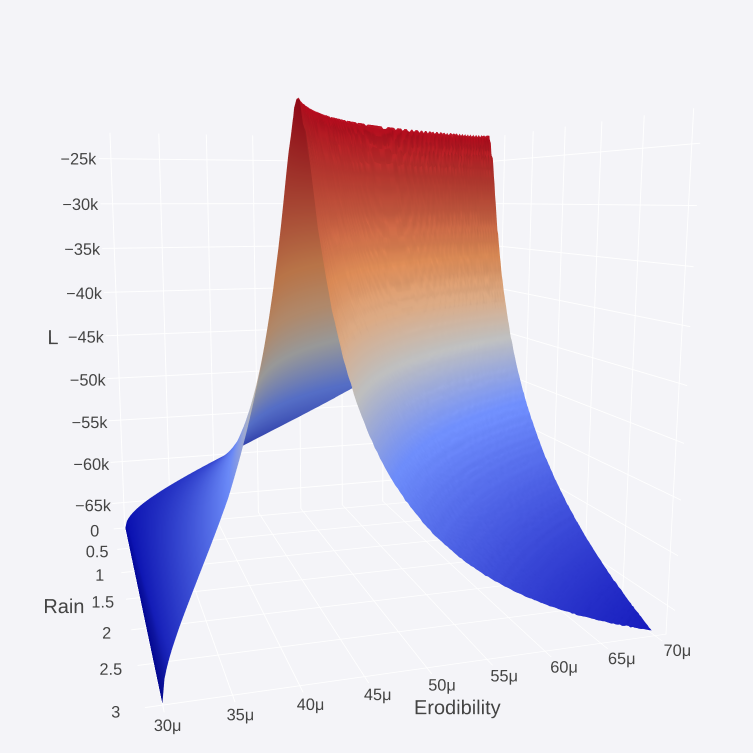}}
      \subfigure[CM  true posterior 
]{\includegraphics[width=80mm]{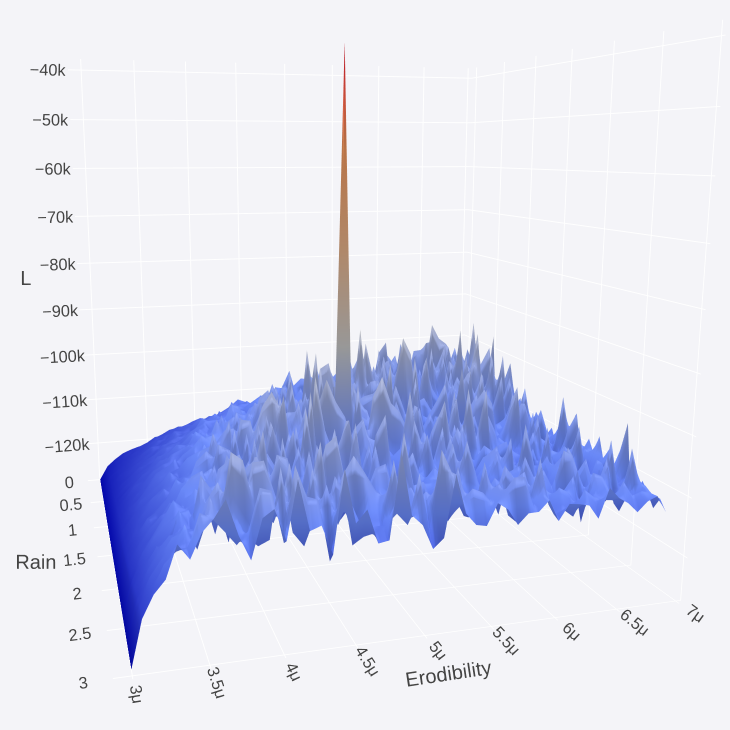}}
    \end{tabular}
    \caption{True  posterior (log-likelihood) surfaces for Cr and 
Continental-Margin topography as a function of precipitation $\rho$ and epsilon 
$\epsilon$ obtained by a grid-search. }
 \label{fig:liklsurface}
  \end{center}
\end{figure*}

To illustrate the difficulty of making inference in geophysical inversion 
problems, we produce the posterior surface (up to a proportionality constant), 
evaluated across a grid of values for $\rho$ and $\epsilon$. We note that it is 
only computationally and visually possible to produce these plots for very low 
dimensional problems, and yet these surfaces  demonstrate the   challenges of 
MCMC sampling. Figure~\ref{fig:liklsurface}~panel~(a) presents the posterior  
surface for the Cr topography  and shows that this surface is more sensitive to 
erodibility  than  precipitation.  It also shows that there is no unique mode 
for precipitation and erodibility. Several combinations of precipitation and 
erodibility give the same posterior value, indicated by the ridge. 
Figure~\ref{fig:liklsurface}~panel~(b) presents  the posterior surface for CM 
which is in sharp contrast to the posterior surface for the Cr. Although it 
shows the existence of a unique global maximum, there are a number of local 
maxima, which makes the  exploration of the surface very challenging.

\begin{figure*}[htbp!]
  \begin{center}
    \begin{tabular}{cc} 
      \subfigure[Cr (true value from Table 
\ref{tab:combinations})]{\includegraphics[width=80mm]{crater/crater_final.png}} 
\\
      
      \subfigure[Cr topography (Combination 1  from Table 
\ref{tab:combinations})]{\includegraphics[width=80mm]{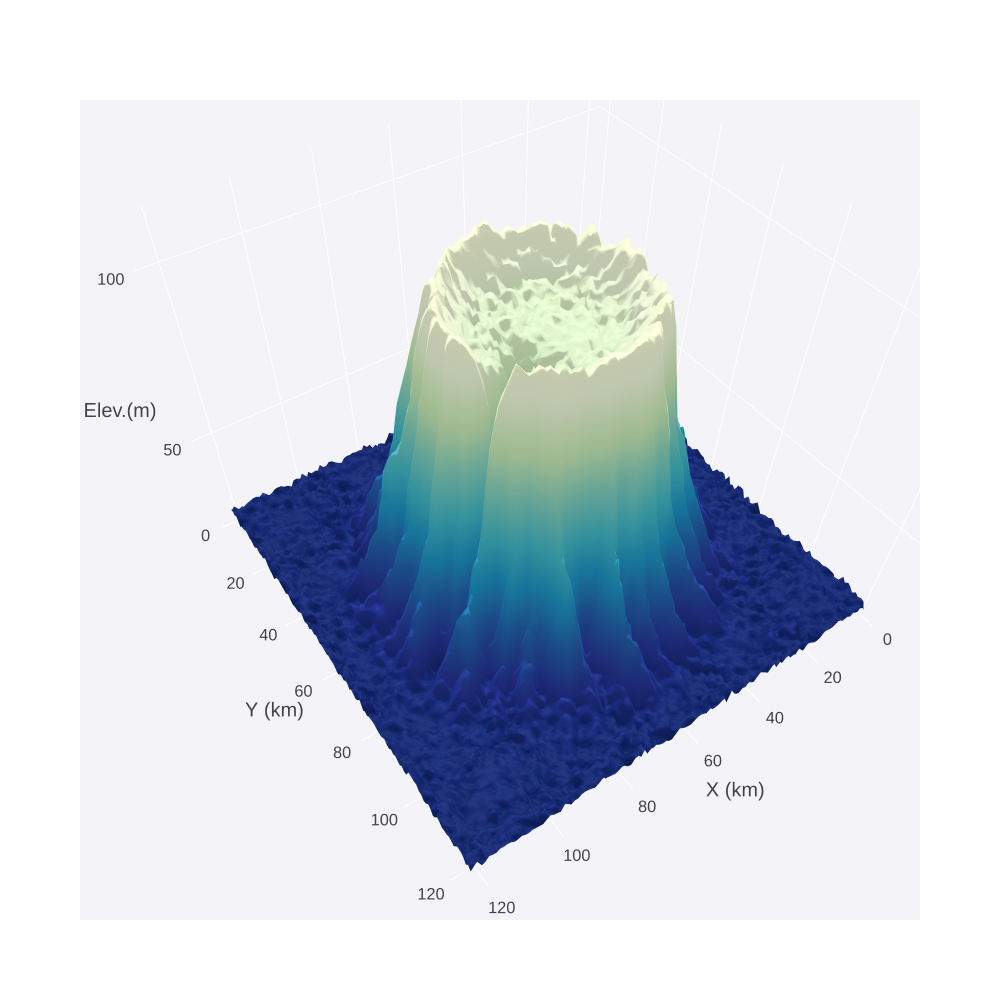}}  
          \subfigure[Cr topography (Combination 5 from Table 
\ref{tab:combinations})]{\includegraphics[width=80mm]{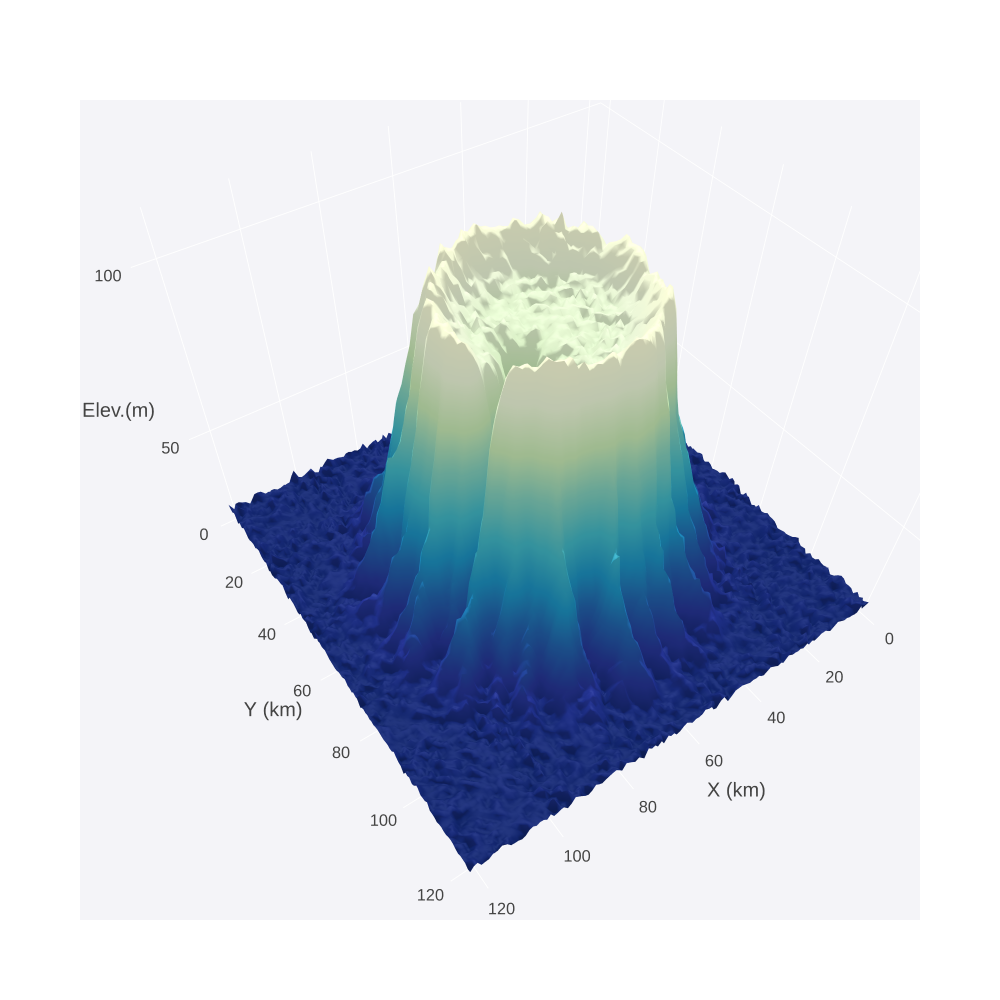}} 
    \end{tabular}
    \caption{Cr: Selected cases showing that there is not much difference 
between the different combinations (unique modes) of the parameters when 
compared to the true-values.  }
 \label{fig:compare_crater-grid}
  \end{center}
\end{figure*}

\begin{figure*}[htbp!]
  \begin{center}
    \begin{tabular}{cc} 
      \subfigure[CM (true 
value)]{\includegraphics[width=80mm]{etopo/etopo_final.png}} \\
      
      \subfigure[CM topography (Combination 
1)]{\includegraphics[width=80mm]{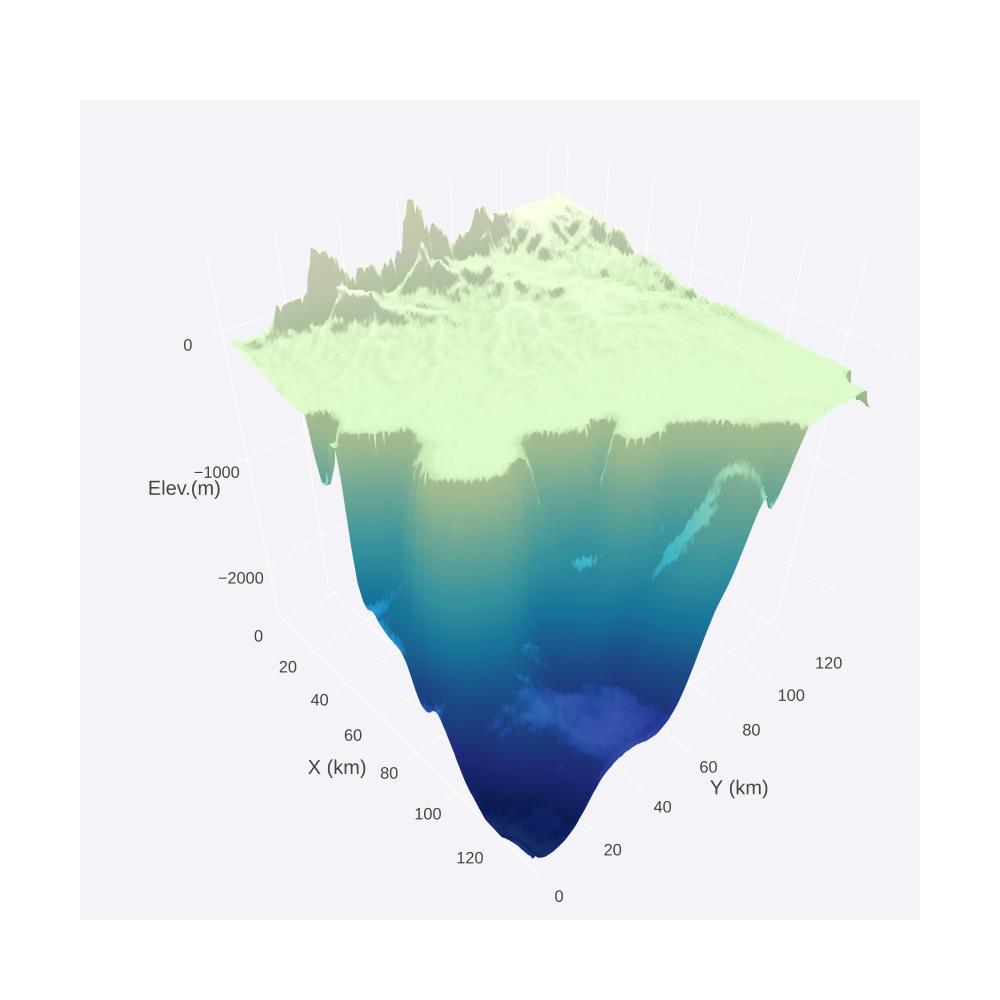}}  
          \subfigure[CM topography (Combination 
5)]{\includegraphics[width=80mm]{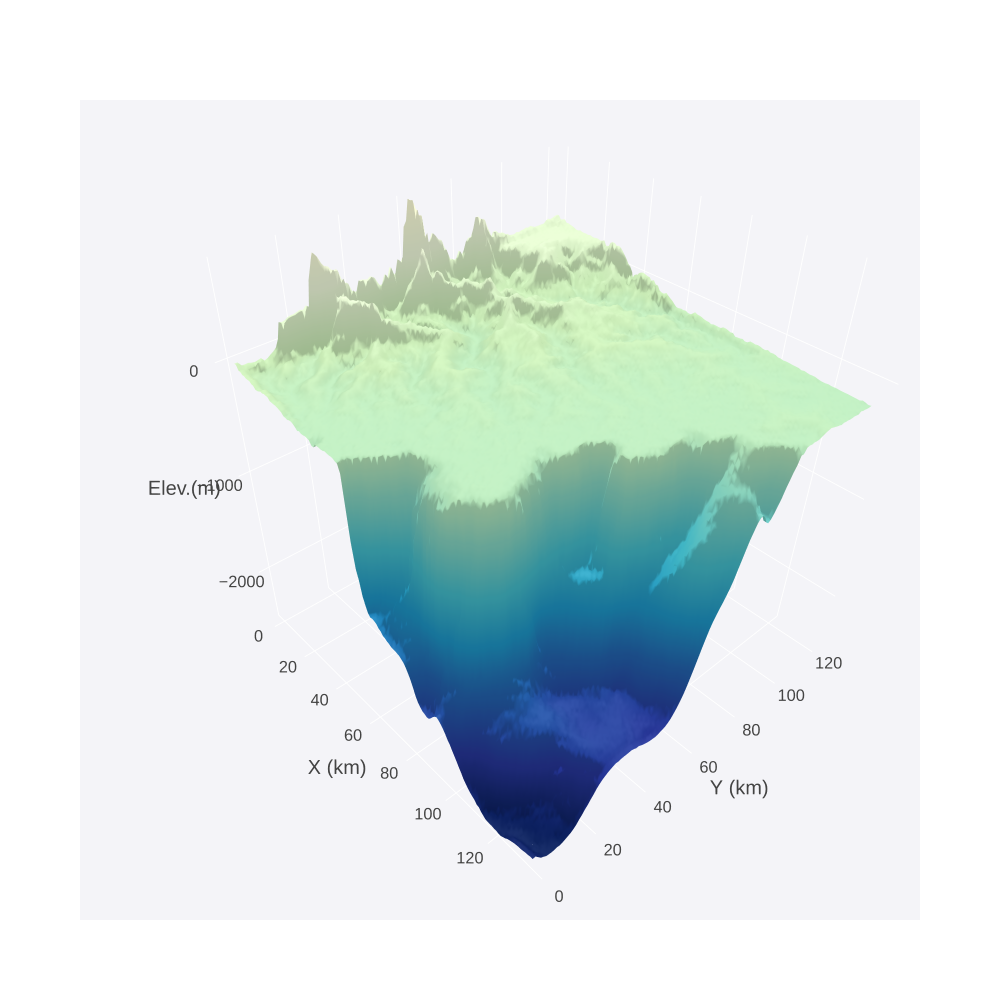}} 
    \end{tabular}
    \caption{CM: Selected cases showing that there is not much difference 
between the different combinations (sub-optimal modes) of the parameters when 
compared to the true-values.  }
 \label{fig:compare_cm-grid}
  \end{center}
\end{figure*}

 \textcolor{black}{Table \ref{tab:combinations} presents the top 10 combinations 
of the parameters 
with minimal  RMSE values. Note that the first combination (true-value) shows 
the actual  values that were used to generate the ground-truth topography, 
hence 
the corresponding RMSE values are   0.  Figures \ref{fig:compare_crater-grid} 
and \ref{fig:compare_cm-grid}  give a visualization of the ground truth 
topography. Panel~(a) corresponds to the topography giving the lowest RMSE, 
while panel (b) corresponds to the topography with the $5^{th}$ lowest RMSE. It 
can be seen that there  is not much difference between the sub-optimal  and 
optimal values, particularly for the CM problem in Figure 
\ref{fig:compare_cm-grid}.}

 \begin{table*}[htb!]
 \small
\centering
 \begin{tabular}{c c c c c c c} 
 \hline \hline
Combinaton &Topography & Precipitation  & Erodibility   &  $RMSE_{elev}$ &  
$RMSE_{sed}$ & $RMSE_{total}$ \\  
 \hline 
 \hline
  True-value  &Cr & 1.50  & 5.00e-5  & 0.00 & 0.00 & 0.00 \\ 
 1 &  & 1.14 & 5.72e-5 & 1.05  & 0.26  & 1.31   \\ 
 2 &  & 2.94 & 3.56e-5 & 1.05  & 0.26  & 1.32   \\ 
 3& & 2.40 & 3.96e-5 & 1.04  & 0.27  & 1.32  \\ 
 4 &  & 2.58 & 3.80e-5 & 1.05  & 0.26  & 1.32  \\ 
 5&  & 1.98 & 4.36e-5 & 1.04  & 0.27  & 1.32   \\ 
 6&  & 0.78 & 6.92e-5 & 1.06  & 0.26  & 1.32  \\ 
 7& & 2.28 & 4.04e-5 & 1.05  & 0.27  & 1.32  \\ 
 8& & 1.08 & 5.88e-5 & 1.06  & 0.26 & 1.32  \\ 
 9&  & 1.02 & 6.04e-5 & 1.05  & 0.27  & 1.32  \\ 
 10&  & 0.90 & 6.44e-5 & 1.07 & 0.26  & 1.32   \\
 
 \hline

  True-value  &CM & 1.50  & 5.00e-6  & 0.00 & 0.00 & 0.00 \\ 
 1 & & 2.64 & 3.96e-6 & 69.85 & 27.63 & 97.48  \\
 2 &  & 2.70 & 4.20e-6 & 69.71 & 28.55 & 98.26  \\
 3&  & 1.38 & 4.68e-6 & 69.84 & 33.44 & 103.20 \\ 
 4 &  & 2.10 & 4.28e-6 & 68.59 & 42.70 & 111.35 \\ 
 5&  & 2.58 & 3.40e-6 & 73.42 & 42.97 & 116.43 \\ 
 6&  & 2.34 & 4.04e-6 & 102.2 & 15.30 & 117.51 \\
 7&  & 2.94 & 3.88e-6 & 79.58 & 40.55 & 120.10 \\ 
 8&  & 2.76 & 3.64e-6 & 104.61 & 15.64 & 120.35 \\ 
 9&  & 1.20 & 5.56e-6 & 69.45 & 51.14 & 120.51 \\ 
 10& & 0.78 & 6.76e-6 & 105.72 & 27.98 & 123.77  \\
  \hline
 \end{tabular}
 
\caption{Columns 3-4 are combinations of  parameters of precipitation $\rho$ 
and 
erodibility $\epsilon$ that give near optimal performance, in terms of 
$RMSE_total$, for Cr, top panel, and CM bottom panel. }
 \label{tab:combinations} 
\end{table*}

\subsection{Inference using Bayeslands }

Table \ref{results} gives details on the number of iterations (samples) in the 
MCMC sampling scheme, the the corresponding time taken, and the prediction 
performance as measured by $RMSE_{elev}$ and $RMSE_{sed}$ and their total, 
$RMSE_{total}$.  We find that $\mbox{RMSE}_{total}$ decreased (improved) with 
the increase in sample size from 10,000 to 50,000 for the Cr topography. This 
improvement is in both $\mbox{RMSE}_{elev}$ and $\mbox{RMSE}_{sed}$, suggesting 
that the MCMC scheme for this landscape had not converged within the first 
10,000 samples.A similar tend  is seen in the the CM topography.

\begin{table*}[htbp!]
\centering
\small
 \begin{tabular}{ c c c c c c c} 
 \hline\hline
Topography &  Samples &  Accepted \% & Time (hours) &  $\mbox{RMSE}_{elev} $& 
$\mbox{RMSE}_{sed}$   & $\mbox{RMSE}_{total}$ \\ 
 \hline\hline
  Cr & 10,000 & 2.35  & 2.26  &  3.69 & 4.56 & 8.25
 \\ 
    & 25,000 & 0.28 & 5.43 & 3.10 & 4.34 & 7.44 
  \\ 
    & 50,000 & 0.22  & 8.68  & 1.26 &	0.95 &	2.21
 \\  
  CM & 10,000 & 0.47  & 4.20  & 133.90 &	83.28 &	217.20
  \\
    & 25,000 & 0.05  & 10.66  & 131.5 &	62.64 &	194.10
   \\ 
  & 50,000 &   0.04  & 19.29 & 94.26 &	29.69 &	123.90
   \\  
 \hline
 \end{tabular}
\caption{Results for Bayeslands MCMC sampling}
 \label{results} 
\end{table*}

\begin{figure}[htbp!]
  \begin{center}
    \begin{tabular}{cc} 
             \subfigure[Trace-plot of  accepted 
samples]{\includegraphics[width=80mm]{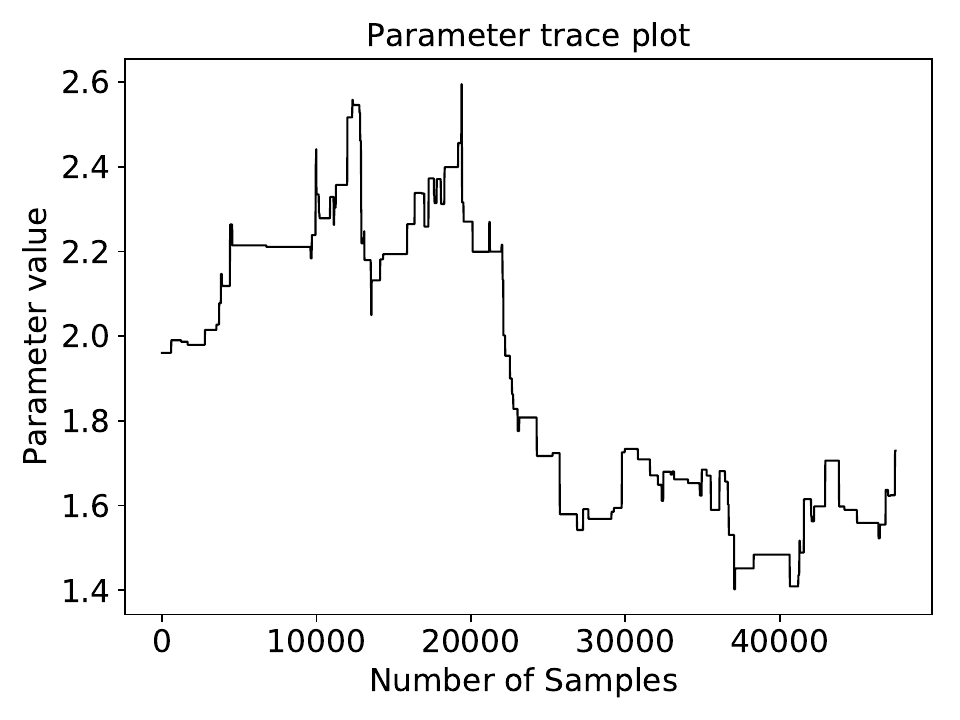}} \\
      \subfigure[Posterior 
distribution]{\includegraphics[width=80mm]{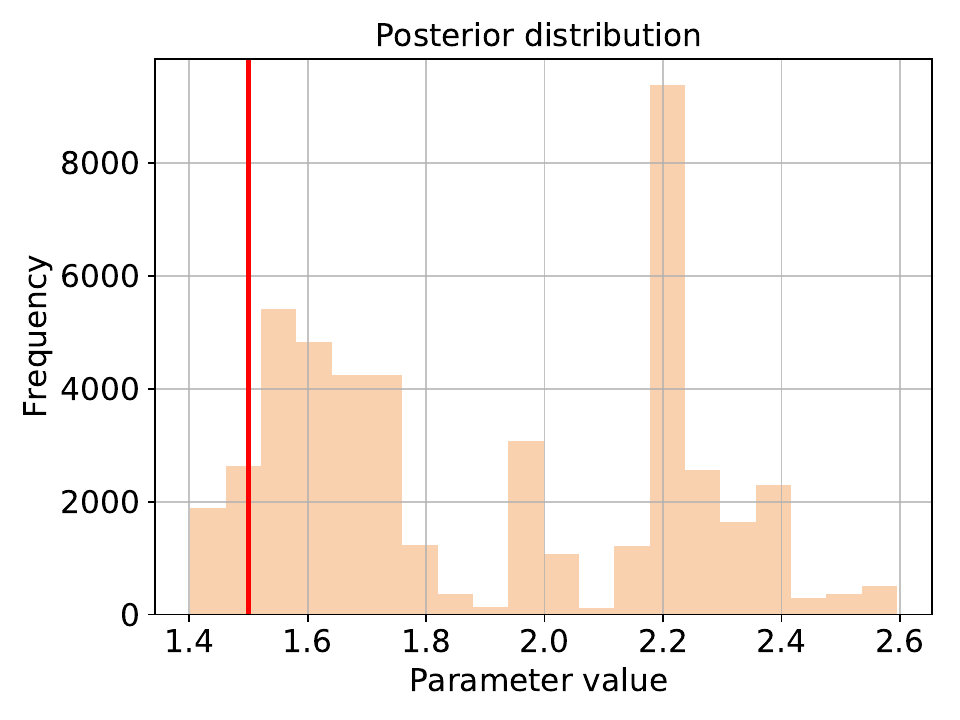}} 
 
    \end{tabular}
    \caption{Cr: Panel~(a) is a trace plot of the Cr precipitation ($\rho$) 
posterior   for  50,000 samples, while panel~(b) is a histogram estimate of the 
posterior distribution. The vertical red line shows true value. }
 \label{fig:crater-pos_rain}
  \end{center}
\end{figure}

\begin{figure}[htbp!]
  \begin{center}
    \begin{tabular}{cc}
    
      \subfigure[Trace-plot of  accepted 
samples]{\includegraphics[width=80mm]{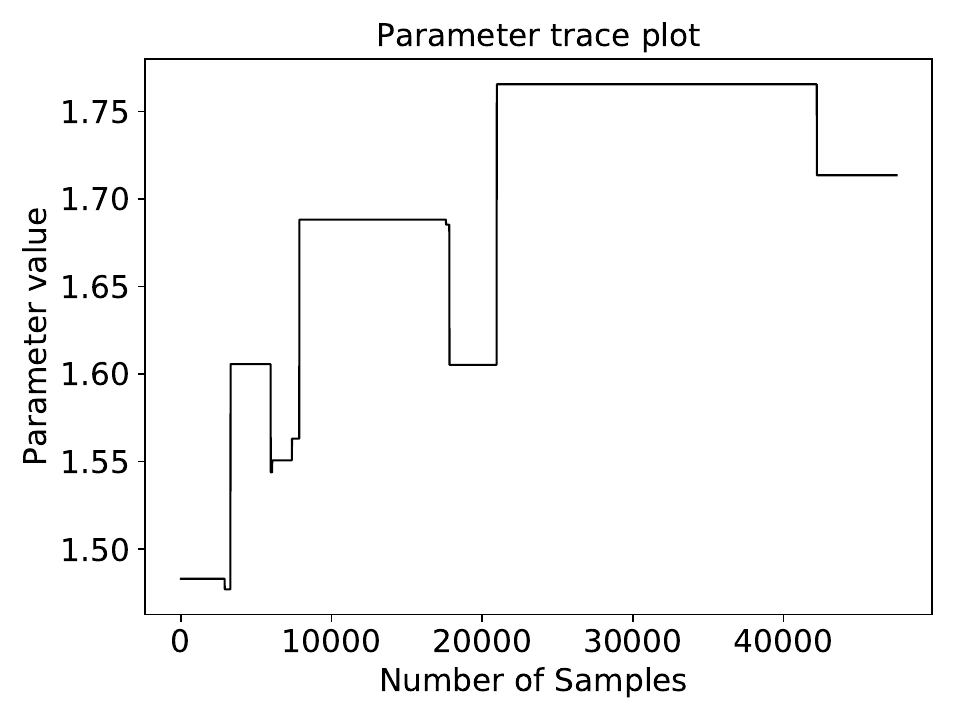}}  \\
      \subfigure[Posterior 
distribution]{\includegraphics[width=80mm]{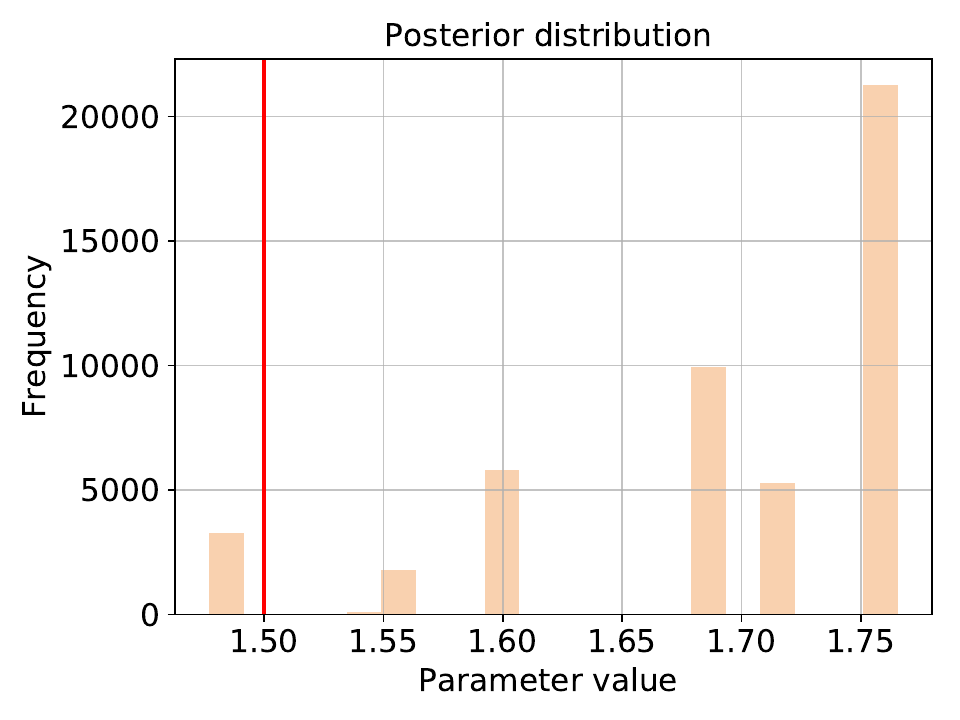}} 
      
    \end{tabular}
    \caption{Panel~(a) is a trace plot of the CM precipitation ($\rho$) 
posterior   for  50,000 samples, while panel~(b) is a histogram estimate of the 
posterior distribution. The vertical red line shows true value. }
 \label{fig:cm-pos_rain}
  \end{center}
\end{figure}

\begin{figure*}[htb!]
  \begin{center}
    \begin{tabular}{cc} 
      \subfigure[12,500 
years]{\includegraphics[width=90mm]{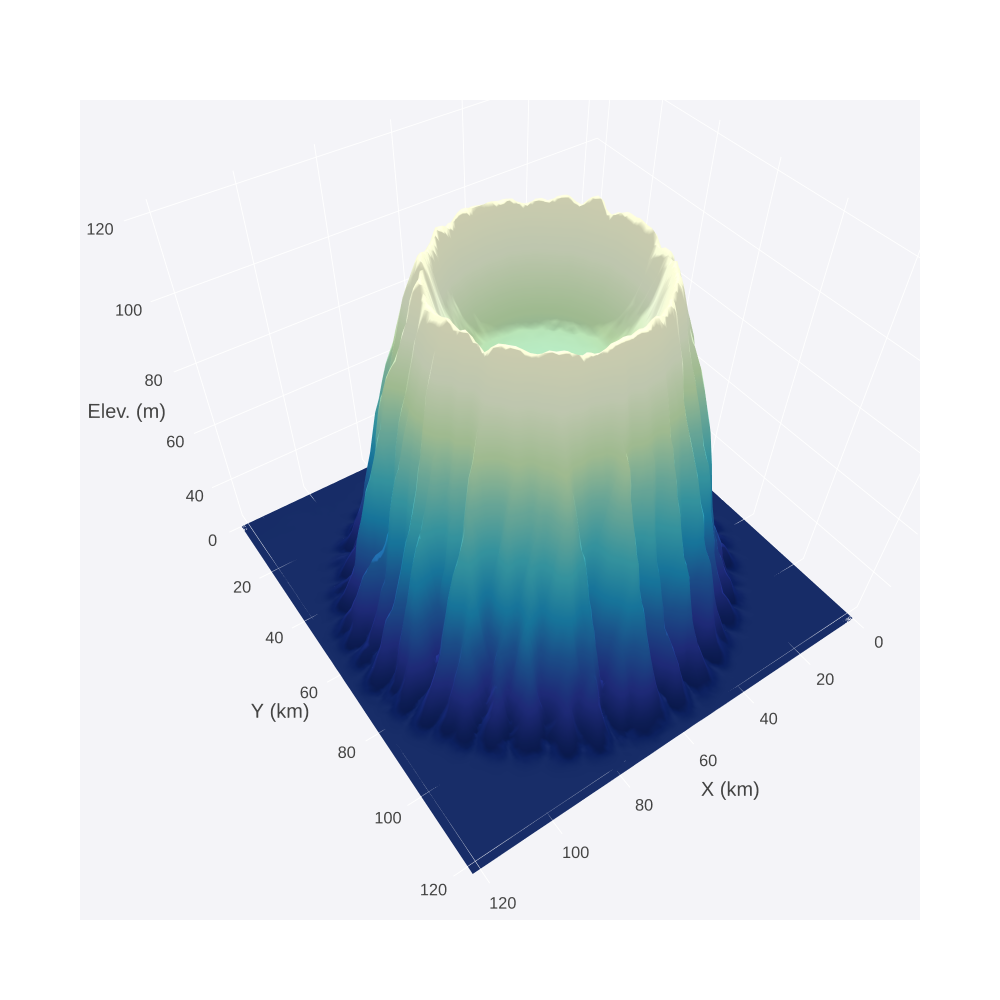}} 
      \subfigure[25,000 
years]{\includegraphics[width=90mm]{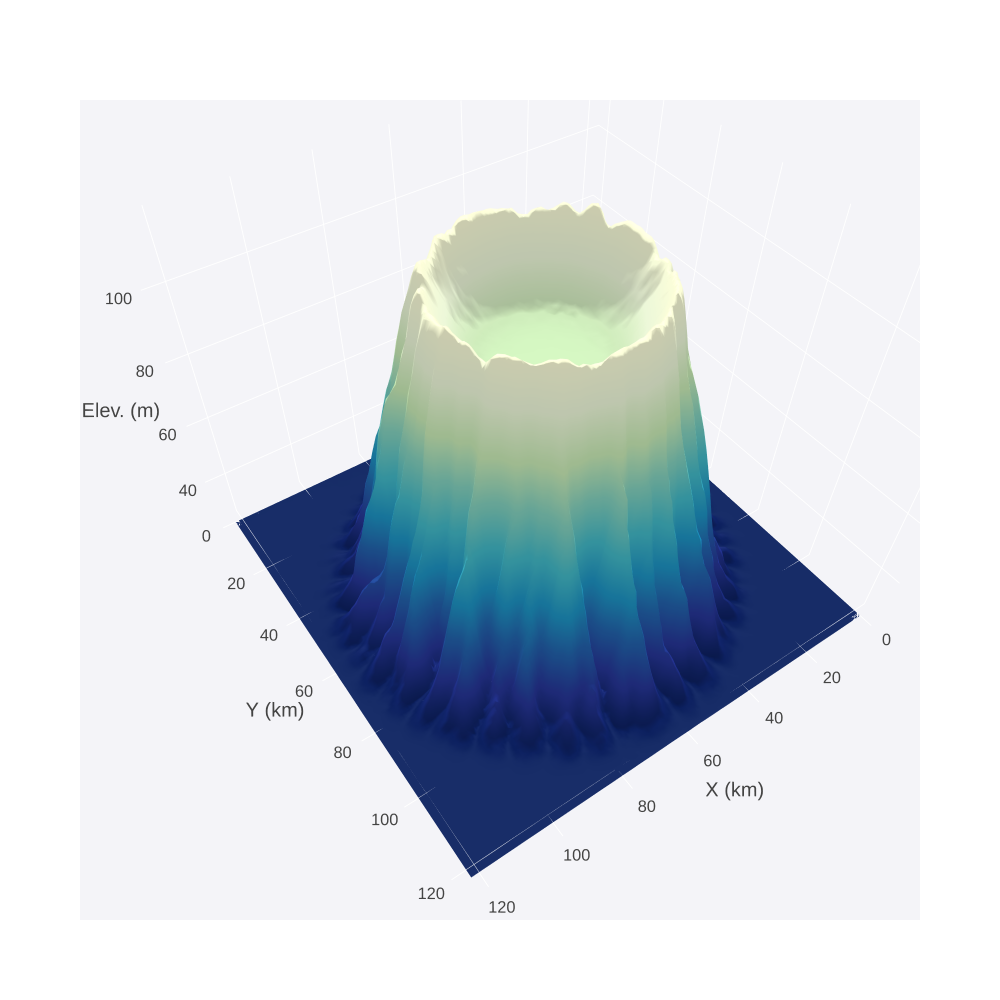}}\\
      \subfigure[37,500 
years]{\includegraphics[width=90mm]{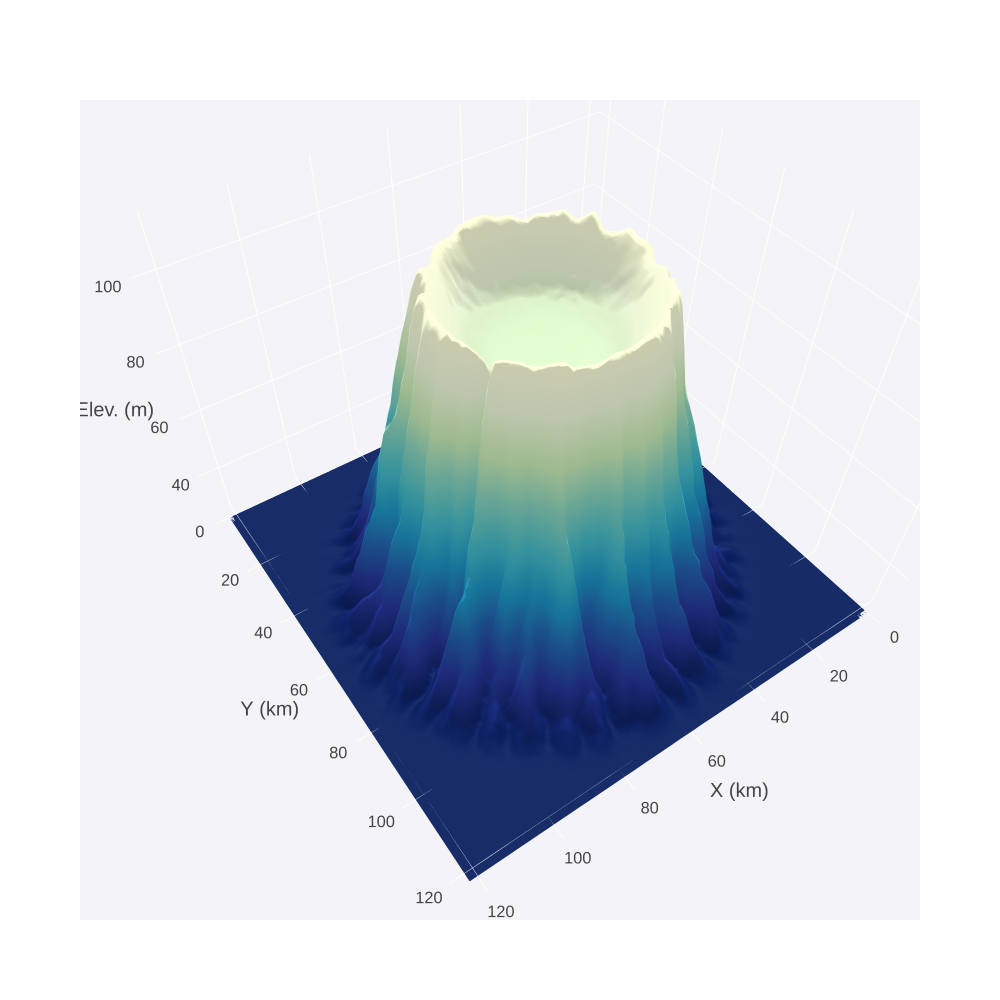}}
      \subfigure[ 50,000 
years]{\includegraphics[width=90mm]{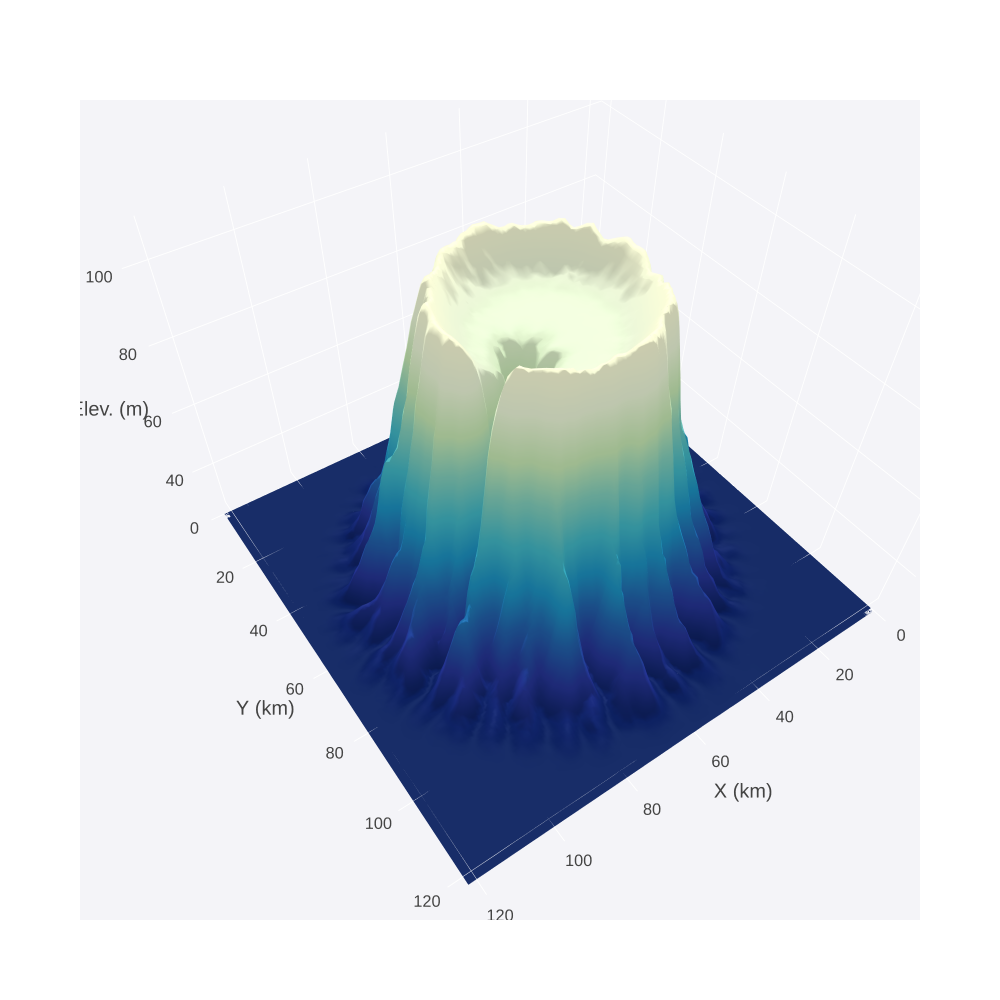}}
    \end{tabular}
    \caption{Cr: Predicted  topographies at selected intervals  for 50,000 
years.}
 \label{fig:crater_topography}
  \end{center}
\end{figure*}

\begin{figure*}[htb!]
  \begin{center}
    \begin{tabular}{cc} 
      \subfigure[250,000 
years]{\includegraphics[width=90mm]{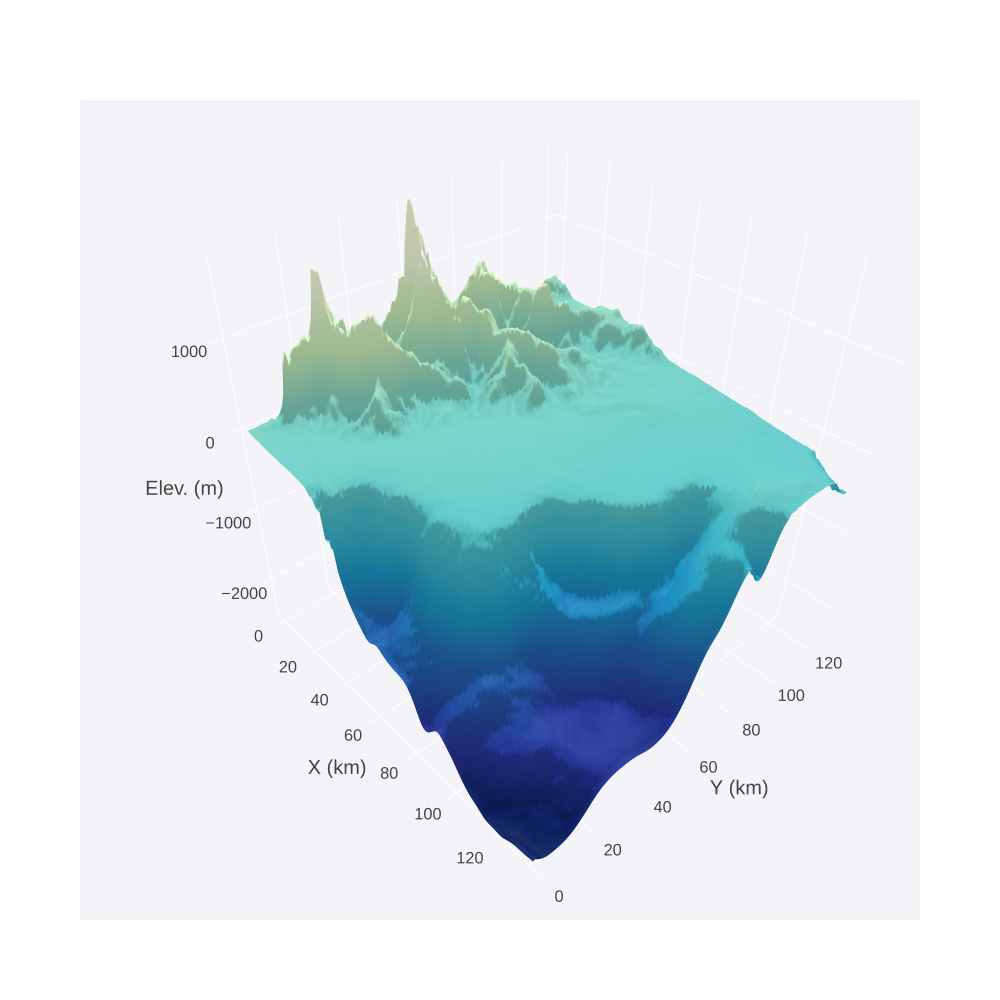}} 
      \subfigure[500,000 
years]{\includegraphics[width=90mm]{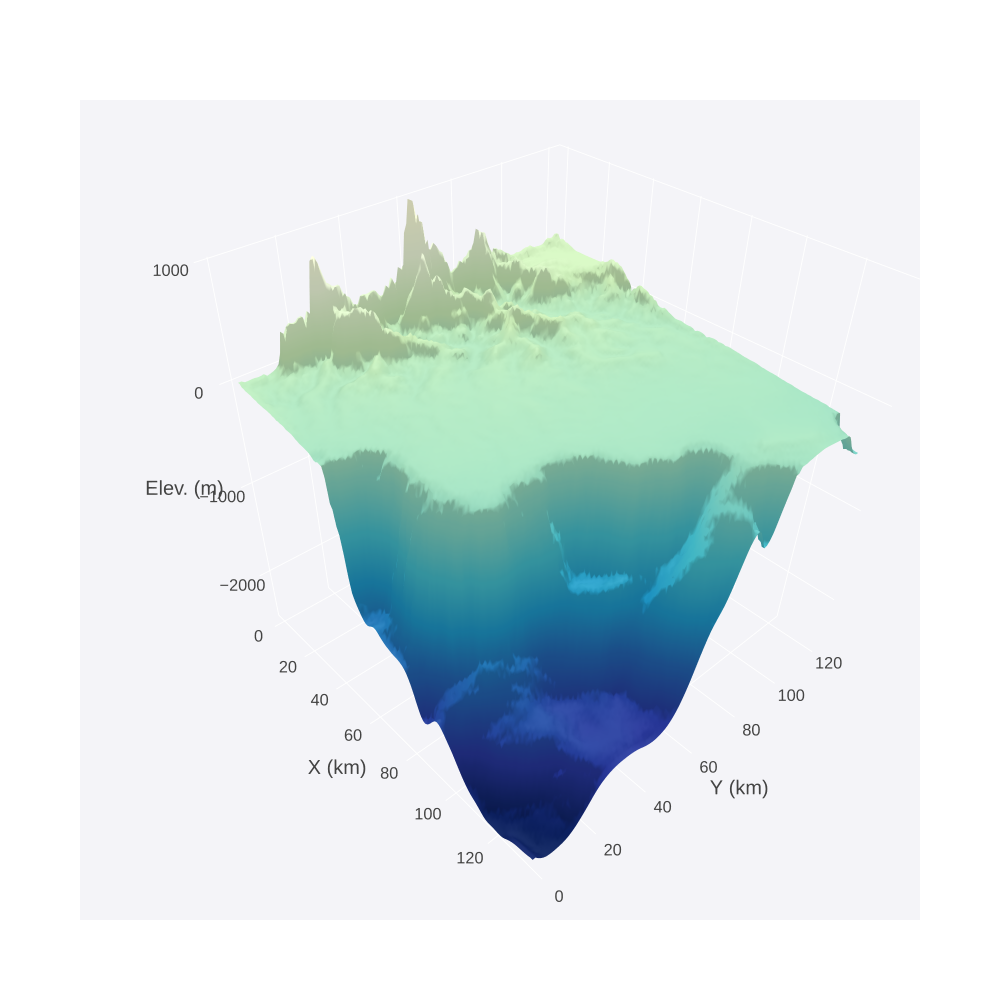}}\\
      \subfigure[ 750,000 
years]{\includegraphics[width=90mm]{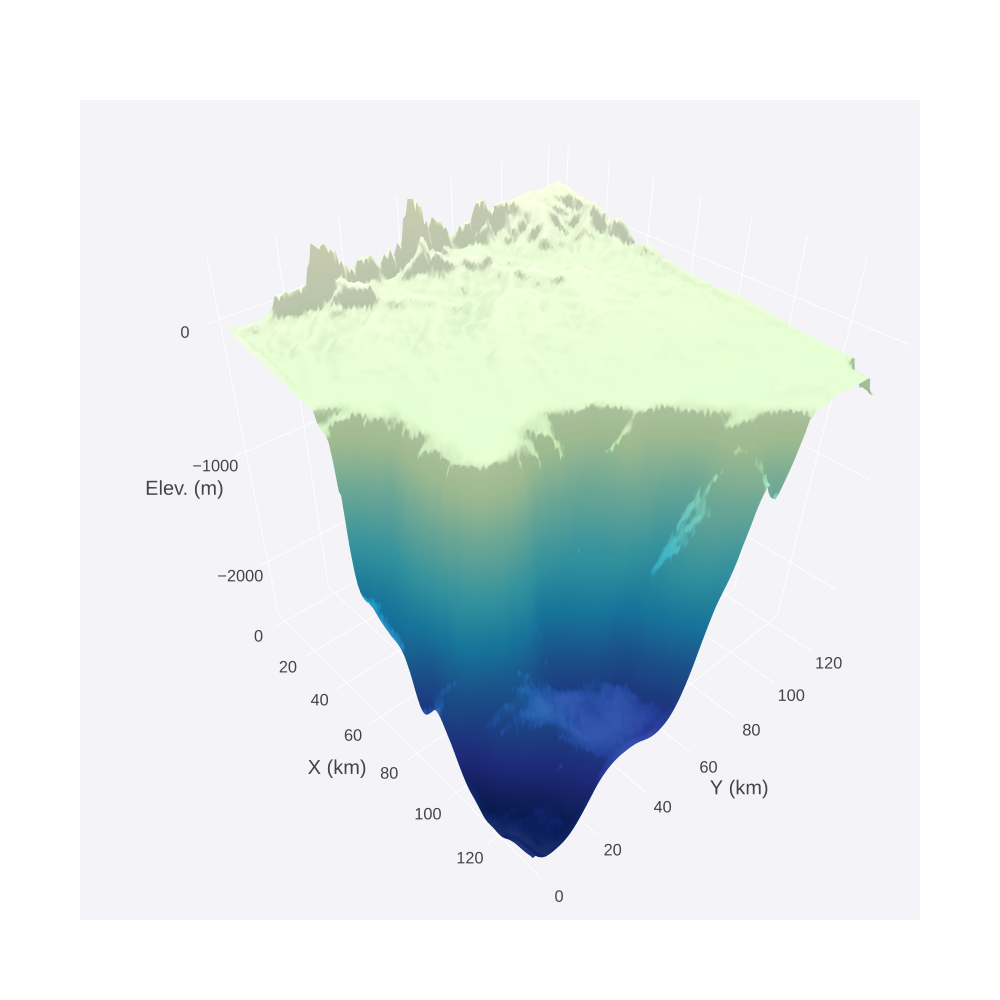}}
      \subfigure[ 1000,000 
years]{\includegraphics[width=90mm]{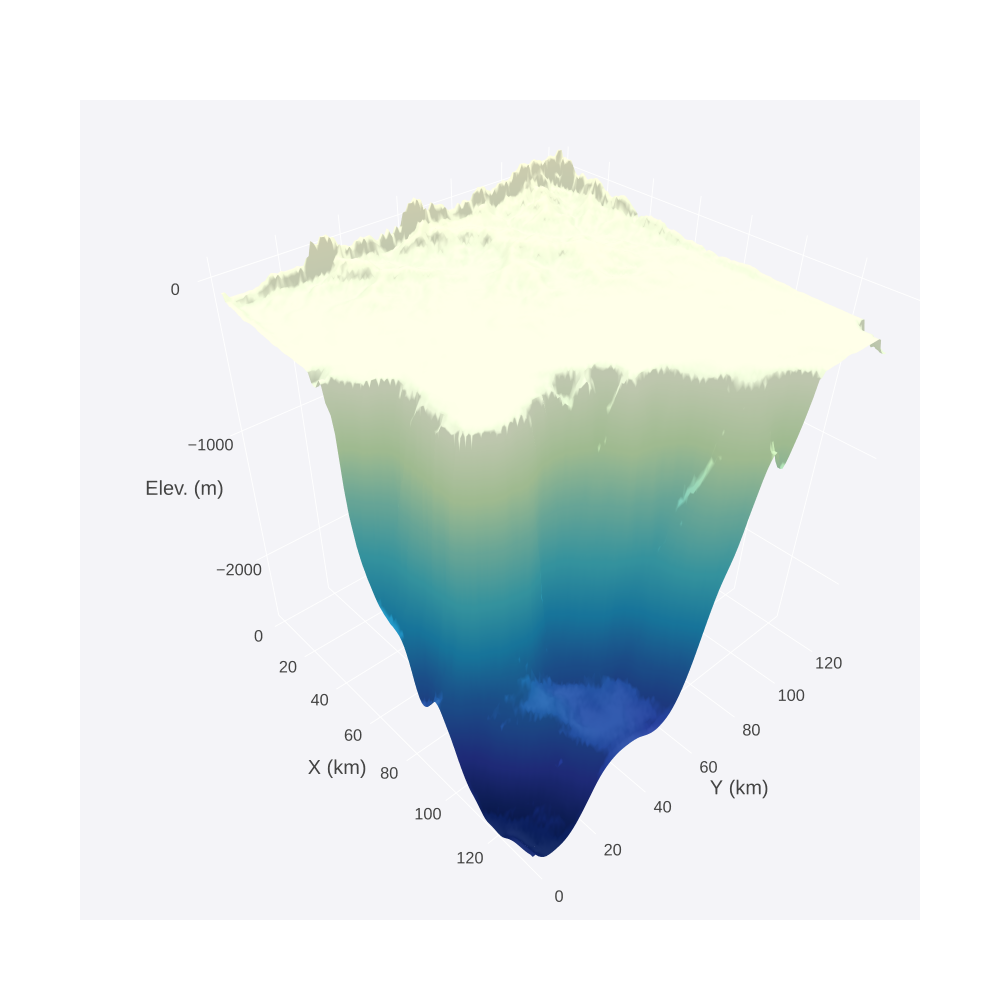}}
    \end{tabular}
    \caption{CM topography evolution over time at selected intervals   for 
1000,000 years.}
 \label{fig:etopo_topography}
  \end{center}
\end{figure*}



\begin{figure*}[htb!]
  \begin{center}
    \begin{tabular}{cc} 
      \subfigure[Elevation cross-section along 
x-axis]{\includegraphics[width=80mm]{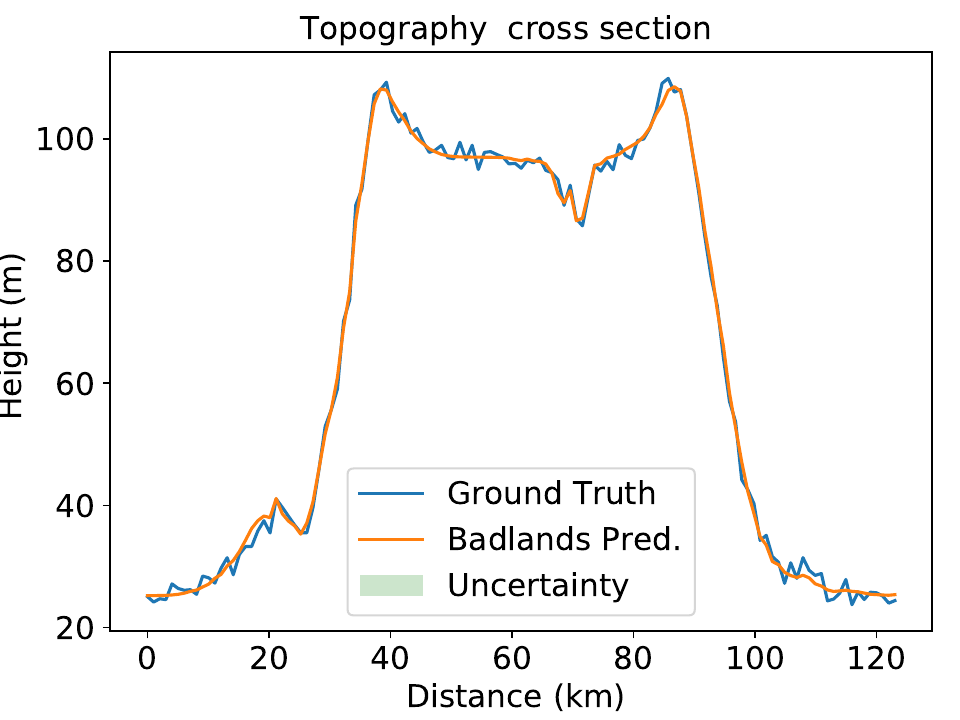}} 
      \subfigure[Elevation cross-section along 
y-axis]{\includegraphics[width=80mm]{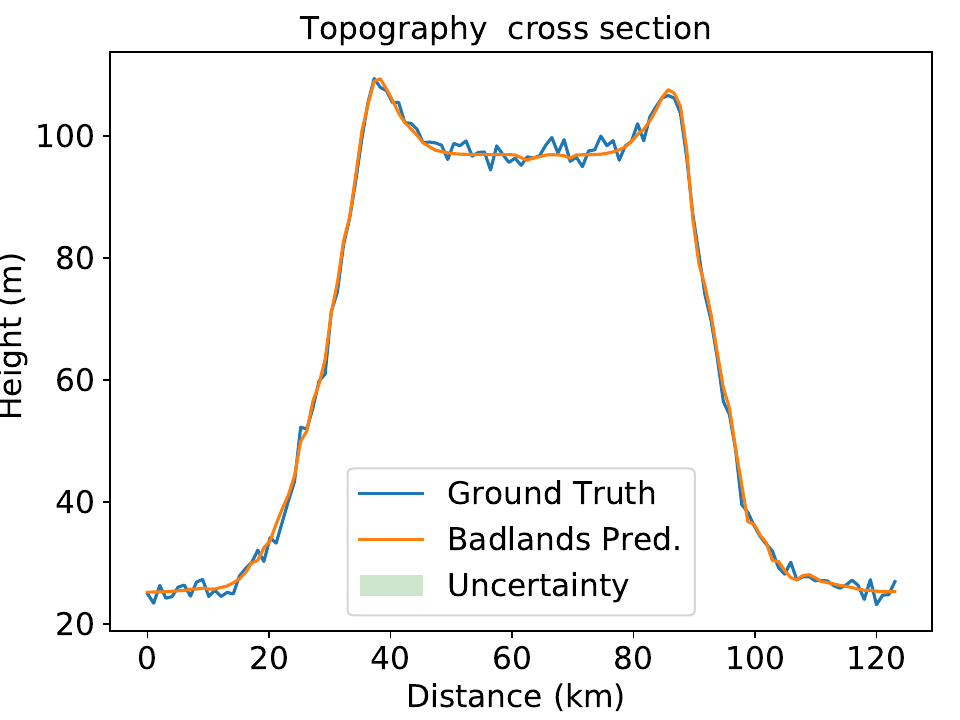}}\\

    \end{tabular}
    \caption{Cr: Elevation cross-section taken at mid-point along x-axis and 
y-axis.The green shaded area corresponds to 95\% credible intervals. }
 \label{fig:crater_cross}
  \end{center}
\end{figure*}

\begin{figure*}[htb!]
  \begin{center}
    \begin{tabular}{cc} 
      \subfigure[Elevation cross-section along 
x-axis]{\includegraphics[width=80mm]{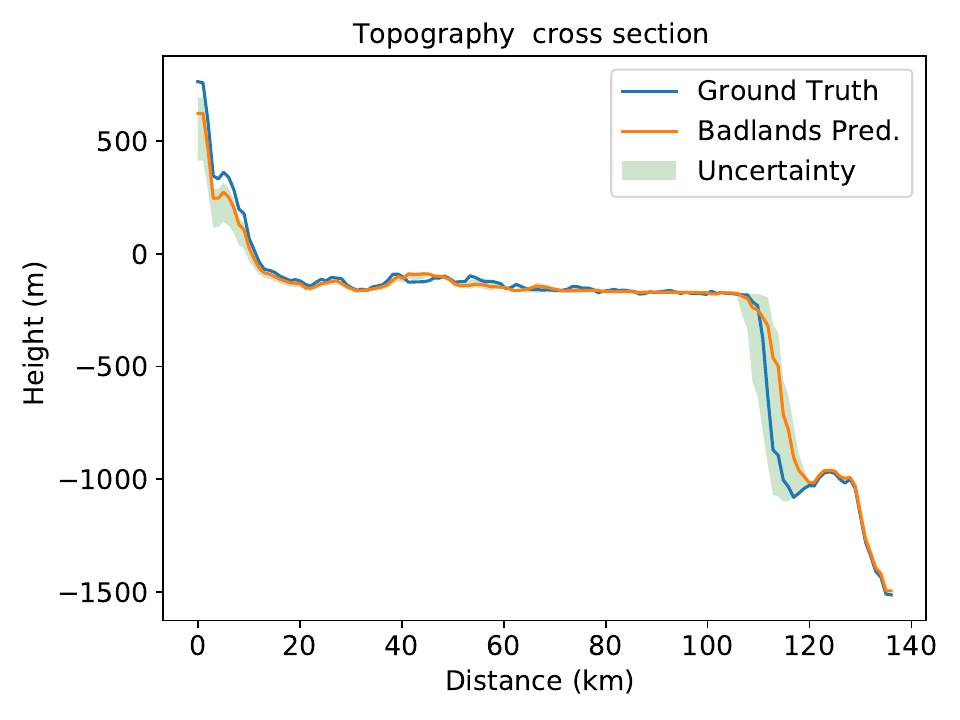}} 
      \subfigure[Elevation cross-section along 
y-axis]{\includegraphics[width=80mm]{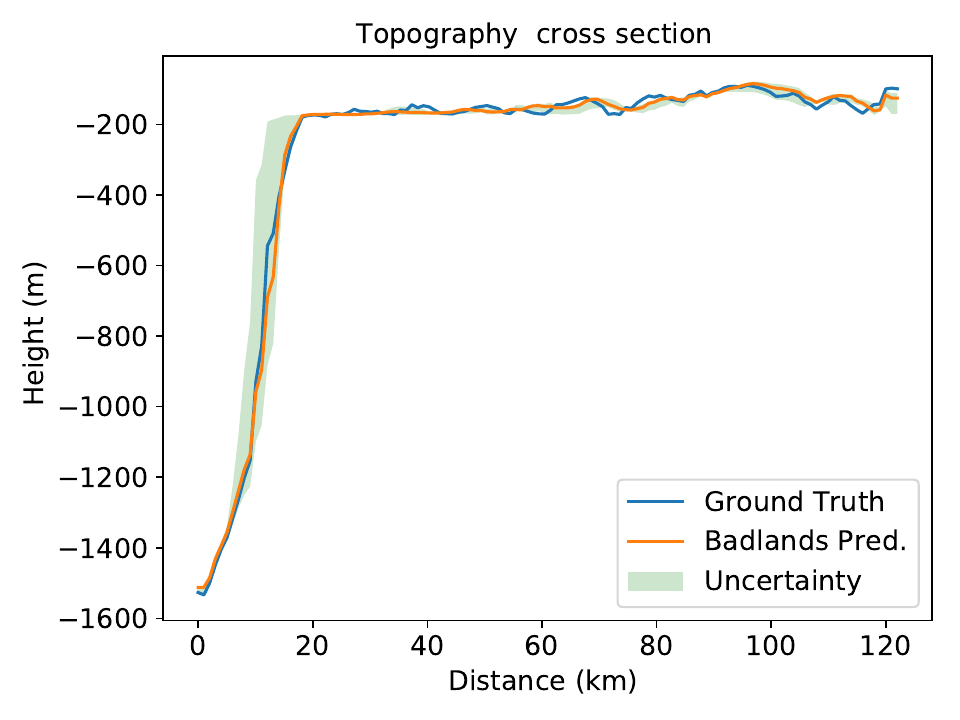}}\\
    \end{tabular}
    \caption{CM model: Panel~(a), elevation cross-section from West to East at 
a 
latitude of $42\degree$. Panel~(b), elevation cross-section from North to South 
at a longitude of 174\textdegree. The green shaded area shows 95\% credible 
interval. }
 \label{fig:etopo_cross}
  \end{center}
\end{figure*}

\begin{figure*}[htb!]
  \begin{center}
    \begin{tabular}{cc} 
      \subfigure[250,000 
years]{\includegraphics[width=75mm]{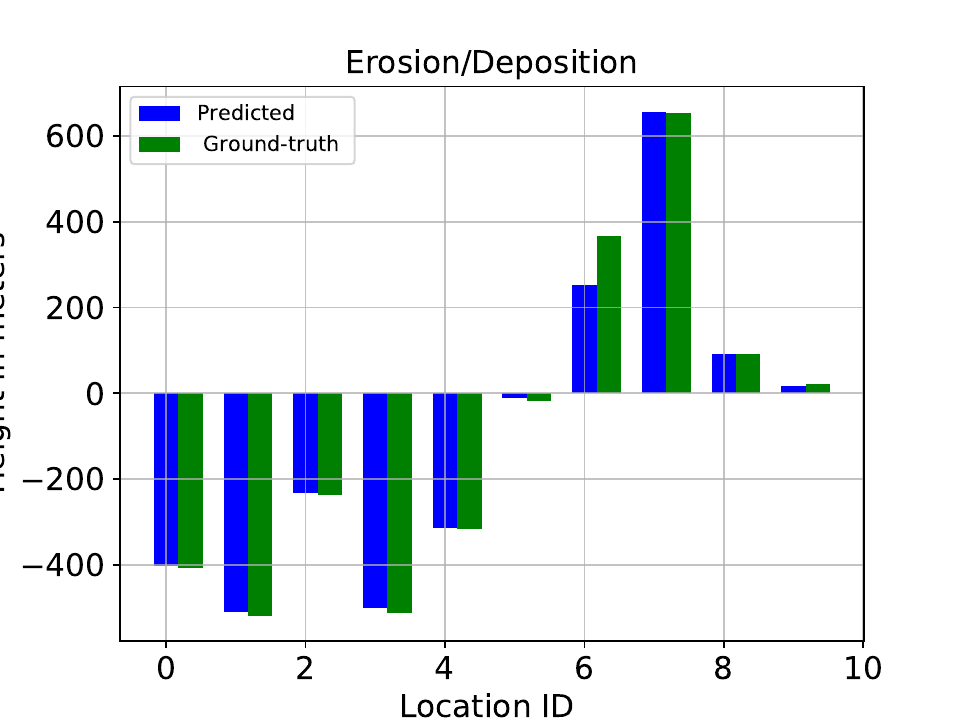}} 
      \subfigure[500,000 
years]{\includegraphics[width=75mm]{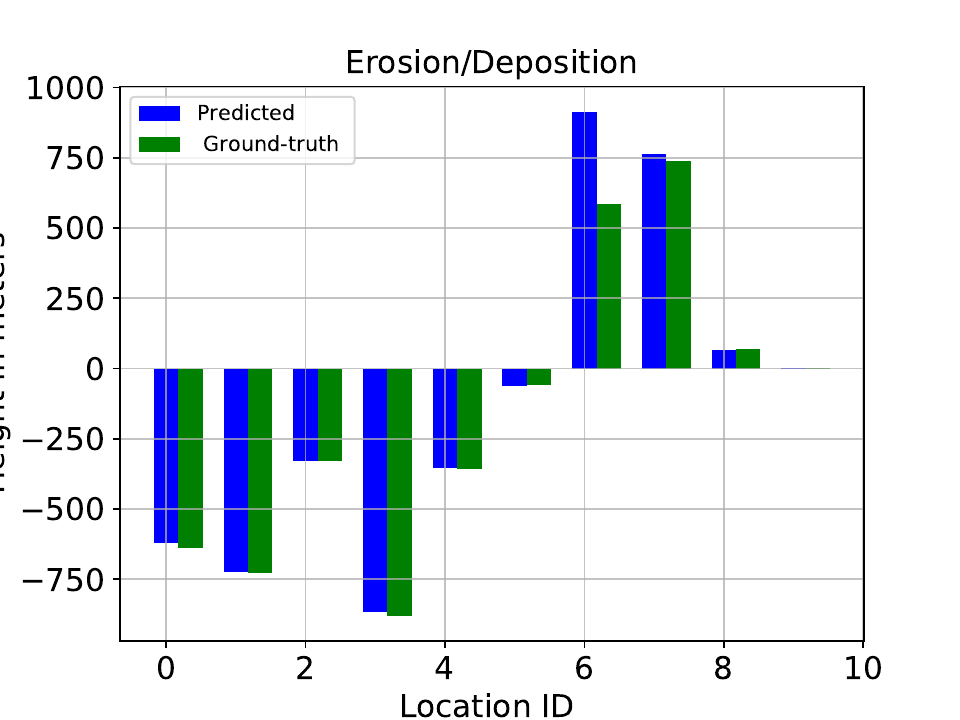}}\\
      \subfigure[ 750,000 
years]{\includegraphics[width=75mm]{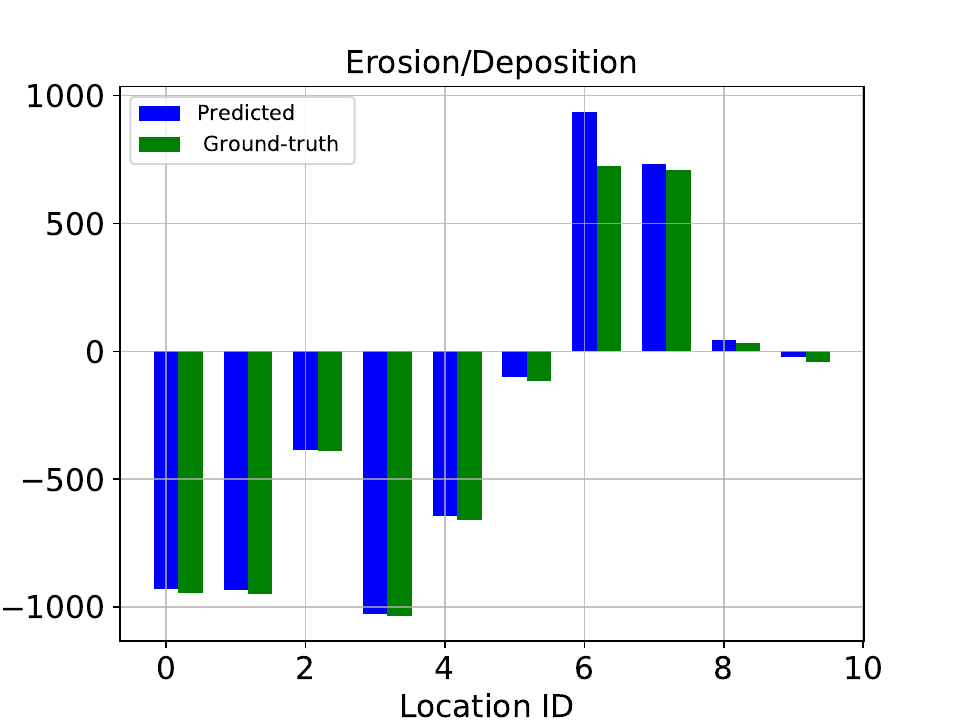}}
      \subfigure[ 1000,000 
years]{\includegraphics[width=75mm]{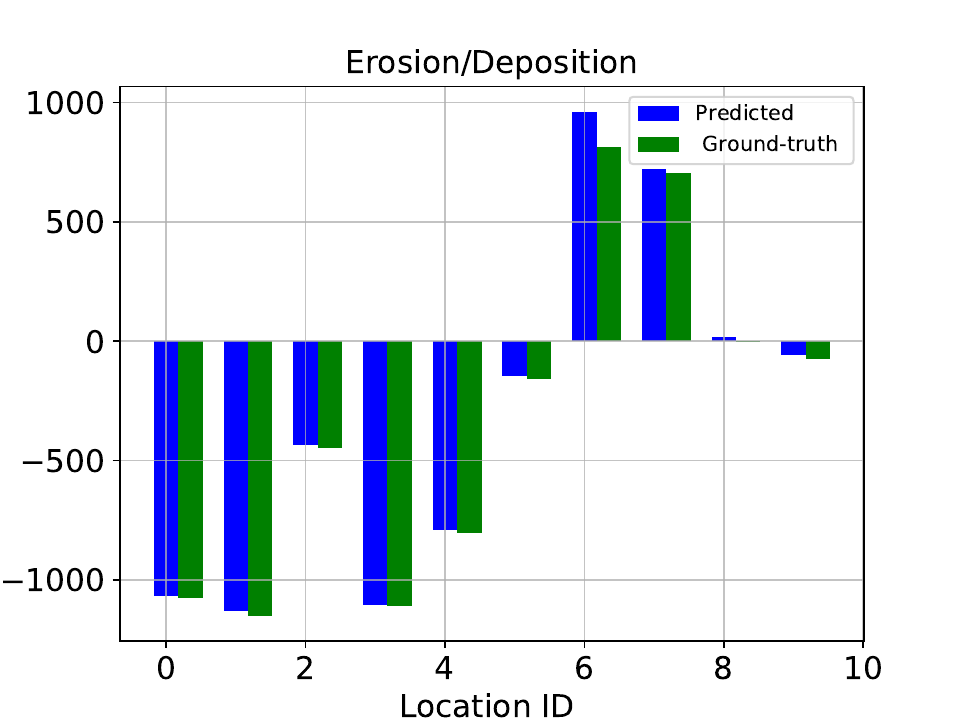}}
    \end{tabular}
    \caption{CM: Sediment erosion/deposition for selected  time-frames fand 
locations for 1,000,000 years. The selected locations (ID) are shown in 
Figure~\ref{fig:erodep_cm}.}
 \label{fig:cm_Sediment}
  \end{center}
\end{figure*}

Figure~\ref{fig:crater-pos_rain},~panel(a) shows the  trace plot of accepted 
proposals in MCMC sampling for precipitation $\rho$,   while panel (b) is a 
histogram estimate of the posterior distribution for the Cr topography.  
Note 
that the vertical line in red in the respective figures shows the true value 
and 
the major goal of the experiments was to check if Bayeslands framework can 
recover the true value. In the case of the Cr topography, we find that  the 
MCMC 
sampling estimate of the posterior distribution is reasonable.  It has one of 
the  modes near the true values of precipitation and erodibility  the MCMC 
chain 
appears to mix well, although with a high degree of auto-correlation. We note 
that the MCMC sampling recovered few different modes as reported by the trace 
plot of the posterior distribution.  In contrast, the CM topography (Figures~ 
\ref{fig:cm-pos_rain}) MCMC sampling procedure shows 
that the chains have difficulty in escaping local modes, hence they are stuck 
(shown in the trace-plot) for longer time periods when compared to the Cr 
topography trace-plots.  As evident by the trace plots, there  little mixing.  
The histogram estimates of precipitation and erodibility are not peaked around 
the true values for these parameters. Instead they are multimodal with modal 
values corresponding to local maxima, given by the coloured dashed vertical 
lines.  The  difference in these results are not surprising, when one considers 
that the CM topography's likelihood surface is very complex when compared with 
that of the Cr, resulting in the need for longer sampling time  and better or 
adaptive MCMC proposal schemes from better traversal.

Despite the difficulty in recovering the  true parameter values, the 
prediction for  the landscape topographies  are astonishingly accurate.  This is evident   from the predicted topography;
Figures~\ref{fig:crater_topography}~and~\ref{fig:etopo_topography} which are 
visually similar to the ground truth, 
Figures~\ref{fig:craterdata}~and~\ref{fig:etopodata}. Moreover,  the 
cross-sections shown in  
Figures~\ref{fig:crater_cross}~and~\ref{fig:etopo_cross} show good accuracy  with  95\% credible 
interval which highlight the accuracy of the prediction. Figure \ref{fig:cm_Sediment} shows the predicted 
evolution of the sediment thickness for both topographies. Interestingly, the 
CM 
predictions  appear to be more accurate than those of the Cr landscape. However 
this may be due to the different scales.  We note that  the sediment thickness 
of the Cr topography is two orders of magnitude less than that  of   CM which  
makes differences appear larger.

 \section{Discussion}

 \textcolor{black}{In this paper, we presented a logically consistent Bayesian 
framework   to estimate and quantify  uncertainty  for selected  parameters 
that 
govern LEMs.  This was achieved by embedding a LEM (Badlands) within a 
probabilistic model.  Although we used Badlands, we stress that the approach is 
general; it can accommodate any LEM and  other geophysical  forward models 
where 
the outputs are a deterministic function of the inputs, for instance Underworld 
models to explore the geothermal potential of the crust 
\cite{quenette2015underworld}. }
 
\textcolor{black}{We made inference for  only 2 free parameters (precipitation 
and erodibility) 
and assumed that they  were constant over space and time.  Even with this 
simplification,  Figure~\ref{fig:liklsurface} shows that the posterior 
(likelihood) surfaces are non-standard and difficult to estimate.
Multi-modality   \cite{charvin2009bayesian} in the posterior distribution is a 
major concern.  Multi-modality  shows that multiple
 combinations of the given parameters (precipitation and erodibility)  can 
plausibility predict or simulate the same topography,  which closely resembles 
the synthetic or ground-truth topography \cite{cross1999construction} (e.g. 
Figure \ref{fig:compare_cm-grid} and Table \ref{tab:combinations}).  In 
addition 
to the issue of  multi-modality,  the posterior may be discontinuous, by which 
we mean the  derivative does not exist which further makes MCMC sampling 
difficult. }
 
 Which aspects of the problem give rise to these unusual posterior surfaces? 
Are 
these surfaces a product of the data or  due to complexity of   the Badlands 
model? It is important  to reflect upon the characteristics or assumptions made 
in geophysical models,   characteristics of the topography data for a given 
time 
and location.   It is worthwhile to investigate if  the posterior distribution 
gives meaningful information  about the appropriateness of the Badlands model 
to 
different topographies.

Clearly, limiting  precipitation and erodibility to be constant for the entire 
spatial and temporal extent of a given model may not be appropriate. While this 
may not make much difference for small geographical regions and short time 
intervals; for larger areas, different regions of the topography  would have 
different distributions of precipitation at different points of time. Moreover, 
the parameter  values are also same throughout the entire topography evolution. 
This does not fully simulate geological time scales and fully capture the  
effects of changing climate through time. In order to implement this,  the 
precipitation would become a vector of parameters that define different regions 
expressed by grids in the map. This  would increase the number of  parameters 
and further add complexity to the model  which would make the inference   more 
difficult.  Furthermore, the synthetic problems should consider larger areas 
that contain a variety of landscape features. This would increase the 
computational complexity of the model and require multi-core implementations of 
MCMC via high performance computing. Parallel tempering 
\cite{earl2005parallel}, 
which is an advanced MCMC method suited for multi-core implementation that 
better captures multi-modality, would be a natural choice to improve the 
performance of the sampler 
\cite{sambridge1999geophysical,sambridge2013parallel}. 

\textcolor{black}{The Bayeslands framework  will form the foundation for more 
complex models of 
landscape and basin evolution. In the future, we envision to include many 
additional parameters in Bayeslands, including the uncertain initial model 
topography,  global sea level fluctuations,  tectonic and dynamic topography 
evolution, spatially varying lithospheric flexural rigidity, spatio-temporal 
variations in mountain uplift rates and in precipitation. Ultimately, we will 
be 
able to consider the uncertain effects of climate change and changing 
vegetation 
cover on spatio-temporal denudation rates and improve assessments on how 
re-vegetation may slow down the continuing erosion of degraded landscapes 
\cite{BASTOLA201825}.}
 
 \section{Conclusions and Future Work}
 
  Bayeslands  provides a framework for   incorporating uncertainty 
quantification for simulated elevation and sediments in Badlands through MCMC 
sampling in landscape evolution models.  Although promising, there are major 
challenges in scaling to higher dimensions which are characteristic of  
real-world applications. As the dimension increases,  there can be further 
challenges with issues of multi-modality and exploration in such  posterior 
distributions.  The exploration of these challenging posterior distributions 
will require the development of new proposal distributions for use in MCMC 
schemes, which reflect local geometry and/or gradient information from the 
Badlands model. 
 
The Bayeslands framework can be  computationally challenging since  Badlands 
takes minutes to hours  for large scale or continental problems. Hence, new 
computationally efficient methods, such as replacing the Badlands model with 
surrogate models for a proportion of the MCMC iterates needs to be developed. 
Speeding up computation of Bayeslands via parallel tempering  in multi-core 
architecture is another avenue to be explored in the future.  
 
 \section{References}
\bibliographystyle{elsarticle-num} 
\bibliography{bibliography,sample} 

\begin{thebibliography}{10}
\bibitem{raje2012bayesian}
D.~Raje, R.~Krishnan, {Bayesian} parameter uncertainty modeling in a macroscale
  hydrologic model and its impact on indian river basin hydrology under climate
  change, Water Resources Research 48~(8).
\end{thebibliography}
 
\end{document}